\documentclass[12pt]{article}
\usepackage{fullpage}
\usepackage{graphicx,times,amsmath,natbib}
\usepackage{amsfonts,amssymb,latexsym,textcomp}
\usepackage{epsfig}
\usepackage{color,xcolor}
\usepackage{subfigure}
\usepackage{url}
\usepackage{theorem}

\newcommand{\bfalpha} {\boldsymbol{\alpha}}
\newcommand{\bfbeta} {\boldsymbol{\beta}}

\newcommand{\bfeta} {\boldsymbol{\eta}}

\newcommand{\bfSigma} {\boldsymbol{\Sigma}}
\newcommand{\bfOmega} {\boldsymbol{\Omega}}

\newcommand{\bfDelta} {\boldsymbol{\Delta}}

\newcommand{\bfz} {\mathbf{z}}

\renewcommand{\Pr}{\mathsf{Pr}}
\newcommand{\E}{\mathsf{E}}
\newcommand{\Var}{\mathsf{Var}}

\newcommand{\normal}{\mathsf{N}}

\newcommand{\IGam}{\mathsf{IGam}}

\newcommand{\bin}{\mathsf{Bin}}

\newcommand{\IWis}{\mathsf{IWis}}

\newcommand{\bet}{\mathsf{Beta}}

\newcommand{\Ber}{\mathsf{Bernoulli}}

\theoremstyle{definition}

\definecolor{gray}{gray}{0.4}

\newcommand\solidrule[1][1cm]{\rule[0.5ex]{#1}{.4pt}}
\newcommand\dashedrule{\mbox{%
  \solidrule[2mm]\hspace{2mm}\solidrule[2mm]\hspace{2mm}\solidrule[2mm]}}

\title{Assessing differences in legislators' revealed preferences:  A case study on the 107$^{\mbox{th}}$ U.S.\ Senate}
\author{
Chelsea L. Lofland\thanks{Graduate Student, Department of Applied Mathematics and Statistics, University of California, Santa Cruz, 1156 High Street, Mailstop SOE2, Santa Cruz, CA, 95064; Fax: (831) 459-4482; Email:  \protect\url{clofland@soe.ucsc.edu}} ,
 \,  
Abel Rodr\'{\i}guez\thanks{Associate Professor, Department of Applied Mathematics and Statistics, University of California, Santa Cruz, 1156 High Street, Mailstop SOE2, Santa Cruz, CA, 95064; Ph.: (831) 459-1047, Fax: (831) 459-4482; Email:  \protect\url{abel@soe.ucsc.edu.edu}  } \, and
Scott Moser\thanks{Assistant Professor, Department of Government, University of Texas at Austin, 1 University Sta. A1800, Austin, TX, 78712; Ph.: (512) 232-7305, Fax: (512) 471-1061; Email: \protect\url{smoser@austin.utexas.edu}   }
}
\date{}
\begin{document}
\maketitle
\begin{abstract}
Roll call data are widely used to assess legislators' preferences and ideology, as well as test theories of legislative behavior.  In particular, roll call data is often used to determine whether the revealed preferences of legislators are affected by outside forces such as party pressure, minority status or procedural rules.  This paper describes a Bayesian hierarchical model that extends existing spatial voting models to test sharp hypotheses about differences in preferences the using posterior probabilities associated with such hypotheses.  We use our model to investigate the effect of the change of party majority status during the 107$^{\mbox{th}}$ U.S.\ Senate on the revealed preferences of senators.  This analysis provides evidence that change in party affiliation might affect the revealed preferences of legislators, but provides no evidence about the effect of majority status on the revealed preferences of legislators.


\end{abstract}

\section{Introduction}\label{se:intro}

Spatial voting models \citep{enelow1984spatial,poole1985spatial,jackman2001multidimensional,clinton2004statistical} are widely used to infer preferences of members of legislative and judicial bodies from their voting records.  {\color{black} Put simply, such models posit that legislators have a most preferred policy -- their \emph{ideal point} --, which can be represented as a point in some Euclidean space, and vote for/against motions in accordance with their (latent) preferences.  Hence, by estimating the ideal point of a legislator from their observed voting behavior we can recover a legislator's \emph{revealed preference} (eg. \citealp{szen06-revealed,rich66-revealed}).  In turn, revealed preferences are often used to construct ideological scales from voting records.  Indeed, while \emph{ideology} has a long a varied history of usage in scholarship, (see eg, Knight 1985), \emph{political ideology} usually refers to specific policy views and preferences held by individuals.  Although political ideology may be based on ``an underlying philosophy on which all specific political views are based'' (pg 17, \citealp{jessee2012ideology}), or stem from logical and psycological sets of constraints influencing a person's beliefs \citep{coverse1964nature}, the notion is often operationalized in terms of revealed preferences and ideal points (e.g., \citealp{pool97-congress}).}  
%

Political scientist often assume that legislators preferences remain stable, at least over short periods of time, (e.g., see \citealp{AJPS:AJPS091}, \citealp{shor2010bridge} and \citealp{PSR:8368174}).  However, although this assumption has some empirical support, the evidence is equivocal.  For example, although \cite{pool97-congress} find that members of congress tend to express very stable policy positions across their careers, \cite{nokken2000dynamics} and \cite{nokken2004congressional} find evidence that legislators who switch party affiliations while serving in Congress exhibit changes in revealed policy preferences, and \cite{jenkins2000examining} and \cite{snyder2000estimating} argue that revealed policy preferences shift across institutional settings and can be influenced by party pressure.  Similarly, \cite{rothenberg2000severing} find evidence that legislators exhibit changes in revealed policy preferences after they have been defeated in a primary or have decided to retire, while \cite{carson2004shirking} make the opposite argument using the same data but a slightly different statistical model.  As a final example, \cite{may1973opinion} and \cite{clausen1987policy} argue that the minority status of a party might affect the revealed preferences of its legislators.  The argument is based on simple game-theoretic models that show the minority party may be able to secure policies closer to their ideal point by staking out extreme positions \citep{merrill1999power}.

The voting record of the 107$^{\mbox{th}}$ U.S.\ Senate is a particularly interesting dataset to investigate questions about the stability of legislators' revealed preferences.  This is because during the 107$^{\mbox{th}}$ U.S.\ Senate, which met between January 3$^{\mbox{rd}}$, 2001 and December 31$^{\mbox{st}}$, 2002, control of the U.S.\ Senate formally shifted within the course of the two-year congressional term.  The 2000 congressional election resulted in a Senate evenly split between Democrats and Republicans.  Because Republican Dick Cheney was elected vice-president in the 2000 presidential election, initially Republicans were considered the majority party, with senator Trent Lott receiving ``the right of first recognition'' and committee chairmanships being assigned to Republican senators\footnote{In spite of Republicans being considered the majority party, Democrats were able to extract some concessions in these unusual circumstances.  For example, committee assignments, staff and other resources were divided equally among the two parties instead of the more usual arrangement in which the majority party receives a bigger share of resources.}.  However, on May 24, 2001, Republican senator James M.\ Jeffords announced that he would leave his party to become an Independent and would caucus with the Democratic Party, putting them in control of the Senate.  This had deep organizational consequences including a transfer in agenda-setting powers to the Democrat leader Tom Daschle and a change in all committee chairmanships.  

\begin{figure}[!h]
\begin{center}
\includegraphics[width=0.90\textwidth,angle=0]{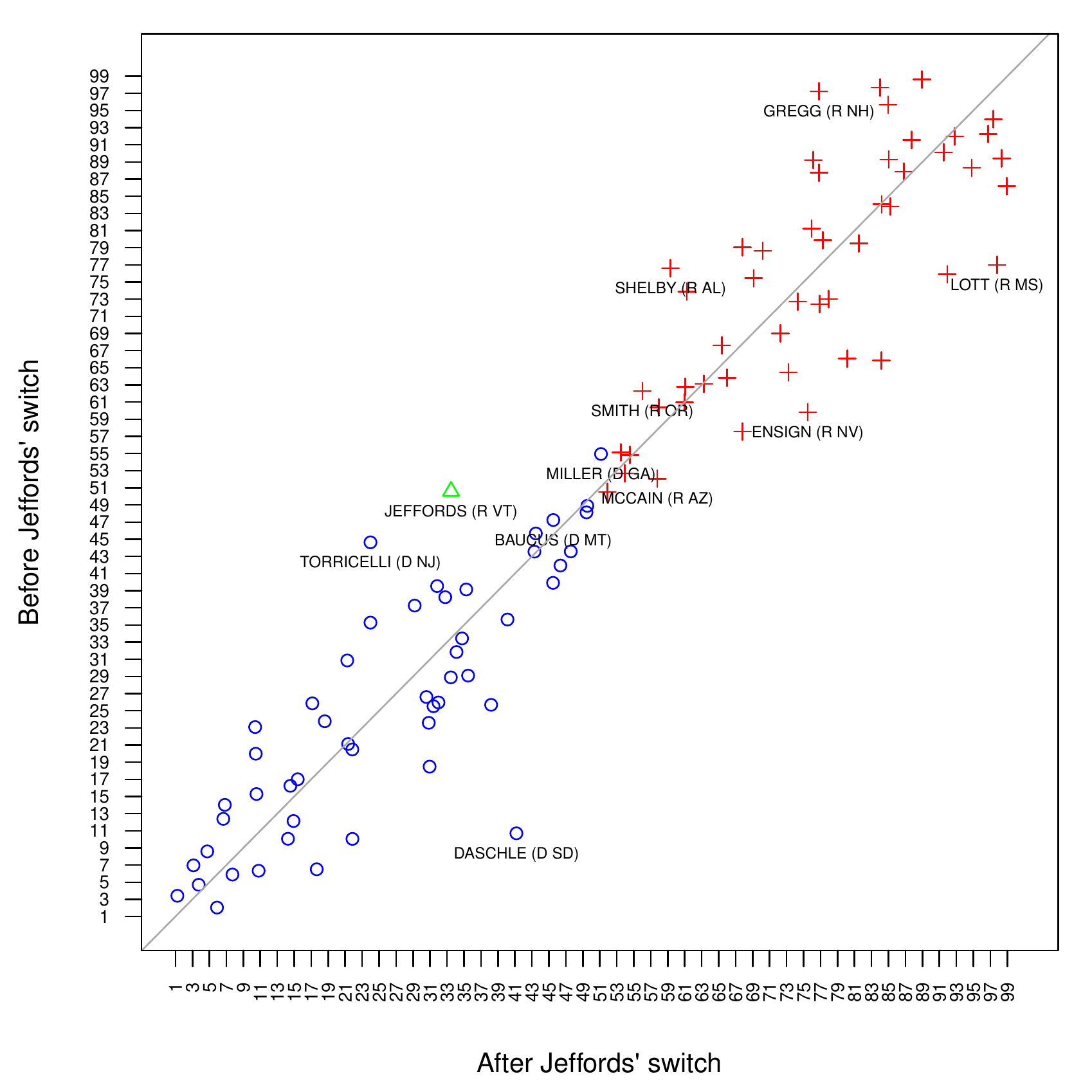}
\caption{Comparison of the ideal point ranks of legislators in the 107$^{\mbox{th}}$ US Senate obtained by fitting independent one-dimensional Bayesian spatial voting models \citep{jackman2001multidimensional} to the motions voted on before and after the defection of senator James M.\ Jeffords from the Republican party.  Democratic senators appear as blue circles ({\color{blue} $\circ$}), Republicans appear as red plus signs ({\color{red} $+$}), and party-switcher Jeffords appears as a green triangle ({\color{green} $\triangle$}).}
\label{fig:independent_ideal_107s}
\end{center}
\end{figure}

\cite{clinton2004statistical} analyzed the roll-call votes from the 107$^{\mbox{th}}$ U.S.\ Senate by fitting separate one-dimensional spatial voting models for motions before and after Jeffords' defection and constructing posterior credible intervals for difference in ideal points as well as for the difference in their rank-order.  This analysis identified a number senators (including Democratic and Republican leaders Daschle and Lott, as well party switcher Jeffords and senators  Gregg, Torricelli, Thompson, Shelby, McConnell, Ensign and Hutchinson) whose preferences seemed to be affected by the change in majority status. On a follow-up analysis, \cite{roberts2007statistical} investigate the apparent change in preferences of the two party leaders (see Figure \ref{fig:independent_ideal_107s}) in terms of their behavior on cloture motions.  In particular, \cite{roberts2007statistical} argues that majority leaders might appear less partisan simply because they sometimes vote strategically against their party during cloture motions so it can be later reconsidered if it fails to receive a qualified majority\footnote{Senate Rule XIII states that ``When a question has been decided by the Senate, any senator voting with the prevailing side ... may, on the same day or on either of the next two days of actual session thereafter, move a reconsideration ... .''}.

Although the conclusions from both of these analyses are certainly plausible and match previous theoretical and empirical work, the methodology used to reach them is unsatisfactory.  One reason is that comparing the rank order of legislators before and after the change in leadership ignores that the ranks of different legislators are not independent.  Indeed, for a legislator to increase her rank, other(s) need to decrease their own.  The obvious alternative is to base the comparisons on the actual ideal points rather than the rank order induced by them.  However, comparing the ideal points directly is difficult because of the invariance of the policy space to affine transformations, and the fact that we have no data linking both policy spaces.  These features imply imply that estimates obtained by fitting separate models to each collection of motions are not directly comparable, as they do not necessarily share a common ideological scale.  \cite{clinton2004statistical} attempt to address this issue by standardizing each set of ideal point estimates to have mean zero and variance.  However, this type of ad-hoc adjustment a posteriori does not really ensure that votes on motions are scaled on a common latent dimension.  A second reason for potential concern with the approaches currently used to assess the stability of legislator's preferences is the lack of adjustment for multiplicities.  Identifying changes in preferences in the U.S.\ Senate implies testing 100 different hypotheses about the relative positions of ideal points, which requires that we adjust our procedures to avoid a large number of false positives.  Although orrections based on Bonferroni (e.g., see \citealp{abdi2007bonferroni}) or False Discovery Rate \citep{benjamini1995controlling} procedures are possible solutions to this type of multiplicity issue, but in our experience they have been rarely used.  The challenge of comparing individual�s traits associated with different latent scales also arises in educational testing, e.g., when attempting to compare the skill level inferred using item response theory (IRT) models \citep{Fo10} for participants in two or more multiple choice tests.  

In this paper we consider a joint model for two groups of motions that allows us to explicitly test sharp hypotheses about differences in legislators' revealed preferences. Our model is an extension of the Bayesian spatial voting models discussed in \cite{jackman2001multidimensional} and \cite{clinton2004statistical} that incorporates hierarchical zero-inflated priors that link the ideal points for the different groups of measures.   Indeed, instead of ad-hoc adjustments to the posterior distribution such as the one proposed in \cite{clinton2004statistical}, {\color{black} our model induces a common scale across both policy spaces by  assuming that not all legislators change their preferences, so that legislators whose preferences remain unchanged serve as a bridge between the two policy spaces \citep{shor2010bridge,PSR:8368174}.}

Although our model is built in a similar spirit, our approach differs from that in \cite{martin2002dynamic}.  Indeed, we are interested in testing sharp hypotheses about changes in revealed preferences rather than model their smooth evolution over longer periods.  Our approach is an alternative to frequentist approaches based on the bootstrap (e.g., see \citealp{lewis2004measuring} and \citealp{carroll2009measuring}) and shares the advantages (and disadvantages) of  Bayesian approaches to variable selection based on Bayes factors/posterior probabilities.  {\color{black} In particular, our simulation studies suggest that our model automatically addresses issues of multiple comparisons, an observation that is consistent with \cite{scott2006exploration} and \cite{scott2010bayes}.}

The remainder of the paper is organized as follows:  Section \ref{se:model} describes the structure of our statistical model.  Section \ref{se:prior_and_comp} discusses hyperparameter elicitation and some general remarks on our computational implementation of the model.
Section \ref{se:illustration} presents our analysis of the data from the 107$^{\mbox{th}}$ U.S.\ Senate, as well as a number of sensitivity and simulation studies that explore the properties of our model in this application. Finally, Section \ref{se:discussion} presents a short discussion and future research directions.

\section{Model description}\label{se:model}

\subsection{Bayesian spatial voting models}\label{se:onesetmodel}

Let $y_{i,j} \in  \{ 0, 1 \}$ encode the vote cast by legislator $i = 1, \ldots, I$ on motion $j = 1, \ldots, J$, with $y_{i,j} = 1$ if the vote corresponds to a ``Yea'' and $y_{i,j} = 0$ if it corresponds to a ``Nay''.  The spatial voting model of \cite{jackman2001multidimensional} and \cite{clinton2004statistical} assumes that legislators make decisions according random quadratic utility functions
\begin{align*}
U_0\left(\bfbeta_i , \bfz_{j,0} \right) &= - \left\| \bfbeta_i - \bfz_{j,0} \right\| ^2 + \epsilon_{j,0}  ,       &
U_1\left(\bfbeta_i , \bfz_{j,1} \right) &= - \left\| \bfbeta_i - \bfz_{j,1} \right\| ^2 + \epsilon_{j,1}  ,      
\end{align*}
where $U_0$ and $U_1$ are, respectively, the utilities associated with a negative and a positive vote, $\bfbeta_i = (\beta_{i,0,1}, \ldots, \beta_{i,0,K})^T$ corresponds to the position of legislator $i$ in a $K$ dimensional Euclidean policy space (his or her ``ideal point''), $\bfz_{j,0}$ and $\bfz_{j,1}$ are, respectively, the positions associated with a negative and a positive vote on motion $j$ in the same policy space, and $\epsilon_{j,0}$ and $\epsilon_{j,1}$ are random shocks. These assumptions lead to a likelihood of the form
\begin{align}\label{eq:likelihoodgen}
y_{i,j} \mid \mu_j, \bfalpha_j,  \bfbeta_{i} \sim \Ber 
\left(  y_{i,j} \, \middle| \, G\left( \mu_j + \bfalpha^{T}_j \bfbeta_{i} \right) \right) ,
\end{align}
where $\mu_j = \bfz_{j,0}^{T}\bfz_{j,0} - \bfz_{j,1}^{T}\bfz_{j,1}$ controls the baseline probability of a positive vote in motion $j$, $\bfalpha_j =  (\alpha_{j,1}, \ldots, \alpha_{j,K})^T = 2\left( \bfz_{j,1} - \bfz_{j,0} \right)$ controls the effect of the ideal points of the legislators on the probability of a positive vote in motion $j$ and $G$ is an appropriate link function.  For example, if $\epsilon_{j,0} - \epsilon_{j,1}$ follows a standard normal distribution then $G$ corresponds to the probit link, so that 
$$
G\left( \mu_j + \bfalpha^{T}_j \bfbeta_{i} \right) = \Phi \left( \mu_j + \bfalpha^{T}_j \bfbeta_{i} \right)  ,
$$
where $\Phi$ is the cumulative distribution of the standard normal distribution, while if $\epsilon_{j,0} - \epsilon_{j,1}$ follows a standard logistic distribution then
$$
G\left(\mu_j + \bfalpha^{T}_j \bfbeta_{i}\right) = \frac{\exp \left\{ \mu_j + \bfalpha^{T}_j \bfbeta_{i} \right\}}
{1 + \exp \left\{ \mu_j + \bfalpha_j^{T} \bfbeta_{i} \right\}}  .
$$
In Section \ref{se:illustration} we consider both types of links and study the sensitivity of the results to this choice.

The model is completed by specifying prior distributions for the model parameters.  It is common to assume that the intercepts $\mu_1, \ldots, \mu_J$ are independent and identically distributed from a normal distribution with unknown mean $\rho_\mu$ and variance $\kappa^2_\mu$, i.e., $\mu_j \mid \rho_\mu, \kappa^2_\mu \sim \normal \left( \mu_j \mid \rho_\mu, \kappa^2_\mu \right)$.  The hyperparameters $\rho_\mu$ and $\kappa^2_\mu$ are in turn assigned independent Gaussian and Inverse-Gamma priors.  Similarly, we use conditionally independent zero-inflated Gaussian priors,
\begin{align*}
\alpha_{j,k} \mid \omega_{\alpha,k}, \kappa^2_\alpha\sim 
\omega_{\alpha,k} \delta_{0}\left(\alpha_{j,k} \right) + (1 - \omega_{\alpha,k}) \normal \left( \alpha_{j,k} \mid 0, \kappa^2_\alpha \right)  ,
\end{align*}
where $\delta_{0}(\cdot)$ denotes the degenerate measure placing probability one at zero (for example, see \citealp{HaCaSc12}), $\kappa^2_\alpha$ is given an Inverse-Gamma hyperprior, and $\omega_{\alpha,k}$ is given a beta prior.  The use of zero-inflated priors for the components of $\bfalpha_j$ allows us to explicitly test the significance of specific dimensions of the policy space.  In particular, note that if $\alpha_{j,k} = 0$ for all $j$ then the $k$-th component of the policy space has no effect on the voting probability and the effective dimension of the policy space is no larger than $K-1$.  In our specification, $\omega_{\alpha,k}$ is the prior probability that the $k$-th dimension of the policy space does not affect the voting behavior of the legislators in motion $j$; for example, the prior probability that dimension $k$ is irrelevant is simply $\omega_{\alpha,k}^{J}$.  Finally, it is common to assume that the ideal points are independently distributed with $\bfbeta_i \mid \bfeta, \bfSigma \sim \normal\left( \bfbeta_i \mid \bfeta, \bfSigma \right)$.

One challenge associated with the interpretation of spatial voting models is the invariance of the policy space to translations, rotations,  reflexions and rescaling.  A common approach to enforce identifiability in these models is to impose constraints on the mean and variance of the ideal points so that $\bfeta = 0$ and $\bfSigma = \mathbf{I}$ (which removes issues related to location and scale) along with constraints on the vectors $\bfalpha_1, \ldots, \bfalpha_K$ so that $\alpha_{j,j} > 0$ (which takes care of reflections) and $\alpha_{j,k} = 0$ if $k>j$ (which addresses invariance to rotations).  In this paper we follow the alternative approach of fixing the position of $K+1$ legislators in policy space \citep{rivers2003identification,clinton2004statistical}.   We carefully choose the legislators whose ideal points are constrained to facilitate the interpretation of the model.  {\color{black} For example, when modeling the U.S.\ Senate using a unidimensional policy space, we ensure identifiability of the parameters by setting the ideal point of two legislators that are clearly in opposite extremes of the political spectrum to $+1$ and $-1$ respectively.  Examples include the leaders of the Republican and Democratic parties, the whips of each party, or the two senators perceived to be the most extreme.  The results are invariant to the identity of the legislators, as long as they are reasonably well separated and on opposite sides.  This choice enables us to interpret the policy space in terms of a liberal-conservative ideology, at least to the extent in which parties are  ideologically opposites.}  

\subsection{Assessing differences in ideal points}\label{se:twosetmodel}

We consider now a joint model for two groups of motions, identified through (known) indicator variables $\gamma_{1}, \ldots, \gamma_{J}$ such that $\gamma_{j} \in \{0,1\}$.  As discussed in the introduction, the goal of this model is to test whether each legislator's voting behavior expresses different revealed preferences in each of these sets of motions.  In our illustration, these two groups of motions correspond to those passed during the periods when either the Republicans or Democrats controlled the 107$^{\mbox{th}}$ Senate, so that $\gamma_j = 0$ if the $j$-th motion was voted upon before senator Jeffords' defection from the Republican party and $\gamma_j = 1$ otherwise.  In this setting it is natural to extend \eqref{eq:likelihoodgen} so that 
\begin{align}\label{eq:likelihoodlgen2}
y_{i,j} \mid \gamma_j, \mu_j, \bfalpha_j, \bfbeta_{i,0}, \bfbeta_{i,1} \sim \Ber 
\left( y_{i,j} \, \middle| G\left( \mu_j + \bfalpha^{T}_j \bfbeta_{i,\gamma_j} \right) \right) ,
\end{align} 
where $\bfbeta_{i,0} = (\beta_{i,0,1}, \ldots, \beta_{i,0,K})^T$ and $\bfbeta_{i,1} = (\beta_{i,0,1}, \ldots, \beta_{i,0,K})^T$ correspond to the (potentially distinct) ideal points of legislator $i$ on each of the two groups of motions.

The likelihood \eqref{eq:likelihoodlgen2} allows (in principle) for different ideal points for each of the two groups of motions.  If we were to assign independent priors to $\bfbeta_{i,0}$ and $\bfbeta_{i,1}$, fitting this joint model would be equivalent to fitting separate (independent) models to each of the two groups.  Instead, we propose a mixture prior that incorporates the possibility that the ideal points are the same,
{\color{black}
\begin{align}\label{eq:prior1}
\bfbeta_{i,0}, \bfbeta_{i,1} \mid \zeta_i, \bfeta, \bfSigma \sim \begin{cases}
\normal\left( \bfbeta_{i,0} \mid \bfeta, \bfSigma \right) \delta_{\bfbeta_{i,0}} \left( \bfbeta_{i,1} \right)   &   \zeta_i = 1   \\
\normal\left( \bfbeta_{i,0} \mid \bfeta, \bfSigma \right) \normal\left( \bfbeta_{i,1} \mid \bfeta, \bfSigma \right)   &   \zeta_i = 0  ,
\end{cases}
\end{align}

The (unknown) auxiliary variables $\zeta_1, \ldots, \zeta_I$ indicate whether the legislators express the same revealed preferences on both groups of motions or not.  In particular, $\zeta_i = 1$ implies that $\bfbeta_{i,0} = \bfbeta_{i,1}$ (so that the $i$-th legislator is a bridge), whereas $\zeta_i = 0$ implies that $\bfbeta_{i,0} \ne \bfbeta_{i,1}$.  Hence, if $\zeta_i = 1$ for all $i$ then our approach is equivalent to fitting a single spatial voting model to all motions.  On the other hand, if $\zeta_i = 0$ for all $i$ then our model fits conditionally independent ideal points for each legislator and group of motions.  Hence, these latent indicators are the key parameters of interest in our analysis.

Because there is no overlap between the measures being voted upon on each period, identification of model requires that $2(K+1)$ constraints be introduced, $K+1$ associated with the set of ideal points $\bfbeta_{1,0}, \ldots, \bfbeta_{I,0}$, and another $K+1$ associated with $\bfbeta_{1,1}, \ldots, \bfbeta_{I,1}$.  While previous authors have imposed independent sets of constraints on each of these two sets of ideal points, we proceed by fixing the position of $K+1$ ideal points and requiring \textit{at least} $K+1$ bridge legislators, whose ideal points remain fixed.  The idea of using legislators whose preferences do not change over time and get to vote on different groups of motions has been exploited by \cite{shor2010bridge} and \cite{PSR:8368174} to compare ideological biases between state legislatures and the U.S.\ Congress.  However, their approach relies on fixing the identity of bridge legislators beforehand, while the key feature of our model is that we aim at identifying the bridges as part of our analysis.  

In order to make inferences on the identity of the bridge legislators while simultaneously enforcing the presence of at least $K+1$ bridges we define a joint prior on the indicators $\zeta_1, \ldots, \zeta_I$ of the form,
\begin{align*}
p(\zeta_1, \ldots, \zeta_I) &= \frac{\Gamma\left( a + \sum_{i=1}^{I} \zeta_i \right)\Gamma\left( b + I - \sum_{i=1}^{I} \zeta_i \right)  } 
{ 1 - \sum_{s'=0}^{K} {I \choose s'}   \Gamma\left( a + s' \right)\Gamma\left( b + I - s' \right)  }    ,    &  \sum_{i=1}^{I} \zeta_i & = K+1, \ldots, I ,
\end{align*}
where $a$ and $b$ are hyperparameters.  To justify this prior, note that it can be rewritten as
\begin{align*}
p(\zeta_1, \ldots, \zeta_I) = p\left(\zeta_1, \ldots, \zeta_I \mid \sum_{i=1}^{I} \zeta_i = s \right) p\left( \sum_{i=1}^{I} \zeta_i = s \right)
\end{align*}
where $p\left(\zeta_1, \ldots, \zeta_I \mid \sum_{i=1}^{I} \zeta_i = s \right) = {I \choose s}^{-1}$ is uniform on all possible subsets of $s$ bridge legislators and 
\begin{align*}
p\left( \sum_{i=1}^{I} \zeta_i = s \right) &=\frac{{I \choose s} \Gamma\left( a + s \right)\Gamma\left( b + I - s \right)  } 
{ 1 - \sum_{s'=0}^{K} {I \choose s'}   \Gamma\left( a + s' \right)\Gamma\left( b + I - s' \right)  }    ,    &  s & = K+1, \ldots, I , 
\end{align*}
corresponds to a truncated Beta-Binomial prior on the number of bridges.  This last prior on the number of bridge legislators can be motivated by noting that the non-truncated version corresponds to the marginal distribution of a hierarchical model where $\sum_{i=1}^{I} \zeta_i \mid \lambda \sim \bin(I, \lambda)$ and $\lambda \sim \bet(a,b)$ (which is the model suggested in \citealp{scott2006exploration} and \citealp{scott2010bayes} to address multiplicity issues).  We  introduce the truncation to ensure the minimum number of bridges required for the model to be identifiable.

The hyperparameters $a$ and $b$ control the prior mean and variance on the number of bridge legislators.  In particular, setting $a=1$ and $b=1$ leads to a uniform distribution distribution on the number of bridges, so that
\begin{align*}
p(\zeta_1, \ldots, \zeta_I \mid a=1, b=1) &= {I \choose \sum_{i=1}^{I} \zeta_i} \frac{1}{I - K}   ,    &   \sum_{i=1}^{I} \zeta_i & = K+1, \ldots, I  .
\end{align*}
Large values of $a/b$ lead to a large expected number of bridges a priori, while large values of $a+b$ lead to lower prior variance.}


Finally, as in the previous Section, we assume $\mu_j \mid \rho_\mu, \kappa^2_\mu \sim \normal \left( \rho_\mu, \kappa^2_\mu \right)$ independently for all $j$, use conditionally independent zero-inflated Gaussian priors, $\alpha_{j,k} \mid \omega_{\alpha,k}, \kappa^2_\alpha\sim \omega_{\alpha,k} \delta_{0}\left(\alpha_{j,k} \right) + (1 - \omega_{\alpha,k}) \normal \left( \alpha_{j,k} \mid 0, \kappa^2_\alpha \right)$, and assign hyperpriors to the unknown parameters $\rho_\mu$, $\kappa^2_\mu$, $\kappa^2_\alpha$ and $\omega_{\alpha,1}, \ldots, \omega_{\alpha,K}$.

{\color{black}
\subsection{Alternative prior specifications}

Alternative specifications for the joint prior $p(\bfbeta_{i,0}, \bfbeta_{i,1} \mid \zeta_i)$ are possible.  For example, one of the referees suggested setting $\bfbeta_{i,1} = \bfbeta_{i,0} + \bfDelta_{i}$ where $\bfbeta_{i,0} \mid \bfeta, \bfSigma \sim \normal(\bfbeta_{i,0} \mid \bfeta, \bfSigma)$ and
$$
\bfDelta_{i} \mid \zeta_i \sim \begin{cases}
\delta_{0}(\bfDelta_{i})  &  \zeta_i = 1  \\
\normal(\bfDelta_{i} \mid \mathbf{0}, \bfOmega)  & \zeta_i = 0,
\end{cases}
$$
which is in spirit of \cite{clinton2004statistical}.  Note that this specification implies 
\begin{align}\label{eq:prior2}
\left( \begin{matrix}
\bfbeta_{i,0}   \\
\bfbeta_{i,1} 
\end{matrix} \right) \mid \zeta_i = 0 \sim \normal
\left(
\left( \begin{matrix}
\bfeta   \\
\bfeta 
\end{matrix} \right) ,
\left( \begin{matrix}
\bfSigma  &  \bfSigma   \\
\bfSigma  &  \bfSigma + \bfOmega
\end{matrix} \right) 
\right)   ,
\end{align}
as opposed to that by implied by \eqref{eq:prior1},
\begin{align}\label{eq:prior3}
\left( \begin{matrix}
\bfbeta_{i,0}   \\
\bfbeta_{i,1} 
\end{matrix} \right) \mid \zeta_i = 0 \sim \normal
\left(
\left( \begin{matrix}
\bfeta   \\
\bfeta 
\end{matrix} \right) ,
\left( \begin{matrix}
\bfSigma  &  \mathbf{0}   \\
\mathbf{0}  &  \bfSigma 
\end{matrix} \right) 
\right)   .
\end{align}

The main differences between these two priors are 1) the marginal variances of $\bfbeta_{0,i}$ and $\bfbeta_{1,i}$ are different from each other in \eqref{eq:prior2} (and in particular, $\bfbeta_{1,i}$ is forced to have a higher variance than $\bfbeta_{0,i}$), and 2) a priori, there is a non-zero correlation between both ideal points in \eqref{eq:prior2}.  For us, the first feature is particularly problematic.  It is not clear to us why we should assume a priori that one set of ideal points has a higher variance than the other, specially if our working assumption is that both sets of ideal points live in the same policy space.  Furthermore, although in the illustration we discuss in Section \ref{se:illustration} there is a natural ordering to the two groups of measures (it would seem natural to center the ideal points after the switch around the ideal points before), that is not the case in other interesting applications.  Hence, our preference for the exchangeable model \eqref{eq:prior3}, which treats both groups of measures identically.  About the second feature of the proposed prior, we note that \eqref{eq:prior3} could be easily extended to include (positive) correlations between both groups of measures when $\zeta_i = 0$.  However, including correlations can lead to identifiability issues:  in the limit (when the value of the correlations approach one) the conditional prior on $\zeta_i = 0$ becomes identical to the prior conditional on $\zeta_i = 1$.  Hence, working with a model that assumes no correlation not only simplifies elicitation, but it can be considered as the most favorable prior to differences (within the class we consider).
}


\section{Hyperpriors and Computation}\label{se:prior_and_comp}

We discuss now the specification of priors for the unknown hyperparameters in our model. For the hyperprior on the intercepts $\mu_1, \ldots, \mu_J$ we let $\rho_\mu \sim \normal(\rho_\mu \mid 0, 1)$ and $\kappa^2_\mu \sim \IGam(\kappa^2_\mu \mid 2, 1)$, where $\IGam$ denotes the inverse Gamma distribution (in this case, with mean 1).  Note that this choice implies that, marginally, $\E \left\{ \mu_j \right\} = 0$ and $\Var\left\{ \mu_j \right\} = 1$, so that if $\bfalpha_j = \mathbf{0}$, then $\Pr\left( y_{i,j} \right)$ has, a priori, a mean of 0 and approximately 95\% probability of falling in the interval $(0.06, 0.94)$, i.e., we do not favor a priori very extreme values for this probability.  Using a similar argument we set the prior on the variance of the parameters $\alpha_{j,k}$ as $\kappa^{2}_\alpha \sim \IGam(2, 1)$ and the priors for the mean and variance of the random effects as $\bfeta \sim \normal(\bfeta \mid \mathbf{0}, \mathbf{I})$ and $\bfSigma \sim \IWis\left( \bfSigma \mid K + 1, \mathbf{I} \right)$, where $\IWis$ denotes the inverse Wishart distribution (in this case, with mean $\mathbf{I}$).  {\color{black} Centering the hyperprior for $\bfeta$ around 0 and the hyperprior for $\bfSigma$ around $\mathbf{I}$ is natural given that the way we defined the identifiability constraints.}  For the prior probability that a positive vote on the $j$-th motion depends on the $k$-th dimension of the policy space we have $\omega_{\alpha,k} \sim \bet\left(\upsilon/K, 1\right)$ which implies that, for a large value of $K$, the probability of a positive vote on any given motion depends a priori on $\upsilon K/\{ K+\upsilon \} \approx \upsilon$ dimensions of the policy space.  Finally, we set $a=1$ and $b=9$.  For the U.S.\ Senate, this choice implies that we expect an average of 11.8 senators that exhibit different preferences on each group of motion.

Posterior inferences on the model parameters are obtained using Markov chain Monte Carlo (MCMC) algorithms \citep{RoCa05}. Given initial values for the parameters, these algorithms successively updates parameters by sampling from their full conditional distributions.  After an appropriate burn-in period, the simulated values are an approximate representation of the target posterior distribution.  {\color{black} In the case of a probit link, sampling can be simplified by introducing auxiliary Gaussian random variables as described by \cite{albert1993bayesian}.  In the case of a logit link we implement a sampler based on P{\'o}lya-Gamma auxiliary random variables along the lines described in \cite{polson2013bayesian}.}

Identifiability is enforced through a parameter expansion approach \citep{liu1998parameter,ghosh2009default}.  At each iteration of the MCMC, the parameters $\mu_j, \bfalpha_j, \bfbeta_{i,\gamma_j}$ are first sampled without any constraint and then the parameters are transformed by applying an appropriate affine transformation.  Details of the computational algorithm are presented in the on-line supplement.

\section{Changes in revealed preferences in the 107$^{\mbox{th}}$ U.S.\ Senate}\label{se:illustration}

In this Section we analyze the voting record of the 107$^{\mbox{th}}$ U.S.\ Senate introduced in Section \ref{se:intro}.  Recall that in this example our two groups of motions correspond to those voted on under Republican (166 motions voted on before May 24, 2001) and Democratic (467 motions voted on after May 24, 2001) control of the Senate.
%
%
We assume that abstentions (i.e., missing values) are ignorable.  In that regard, we note that only 2.58\% of the votes are missing (so abstentions are relatively rare), and tend to be concentrated in a small number of senators and motions.  
Furthermore, our previous work on the use of abstentions in the U.S.\ Congress suggests that strategic behavior is also relatively rare \citep{rodriguez2013modeling}.  We also note that senator Paul Wellstone (Democrat, MN) died in a plane crash October 25, 2002, and was replaced by senator Dean Barkley.  Hence, we follow \cite{roberts2007statistical} and exclude senator's Barkley (who voted in only 14 roll calls) from the analyses.  All results presented below are based on 50,000 iterations of our Markov chain Monte Carlo algorithms, and we monitored convergence using the multi-chain algorithm described in \cite{GeRu92}. To facilitate comparisons with previous analyses of this dataset we fit a one dimensional model ($K=1$) to the data.  We do not consider this last assumption a limitation, as the U.S.\ Senate is widely acknowledged to be unidimensional \citep{pool87-analysis,pool91-patterns,pool97-congress,mccarty2006polarized}.
%
%
\begin{figure}[!h]
  \begin{center}
  \includegraphics[width=0.8\textwidth]{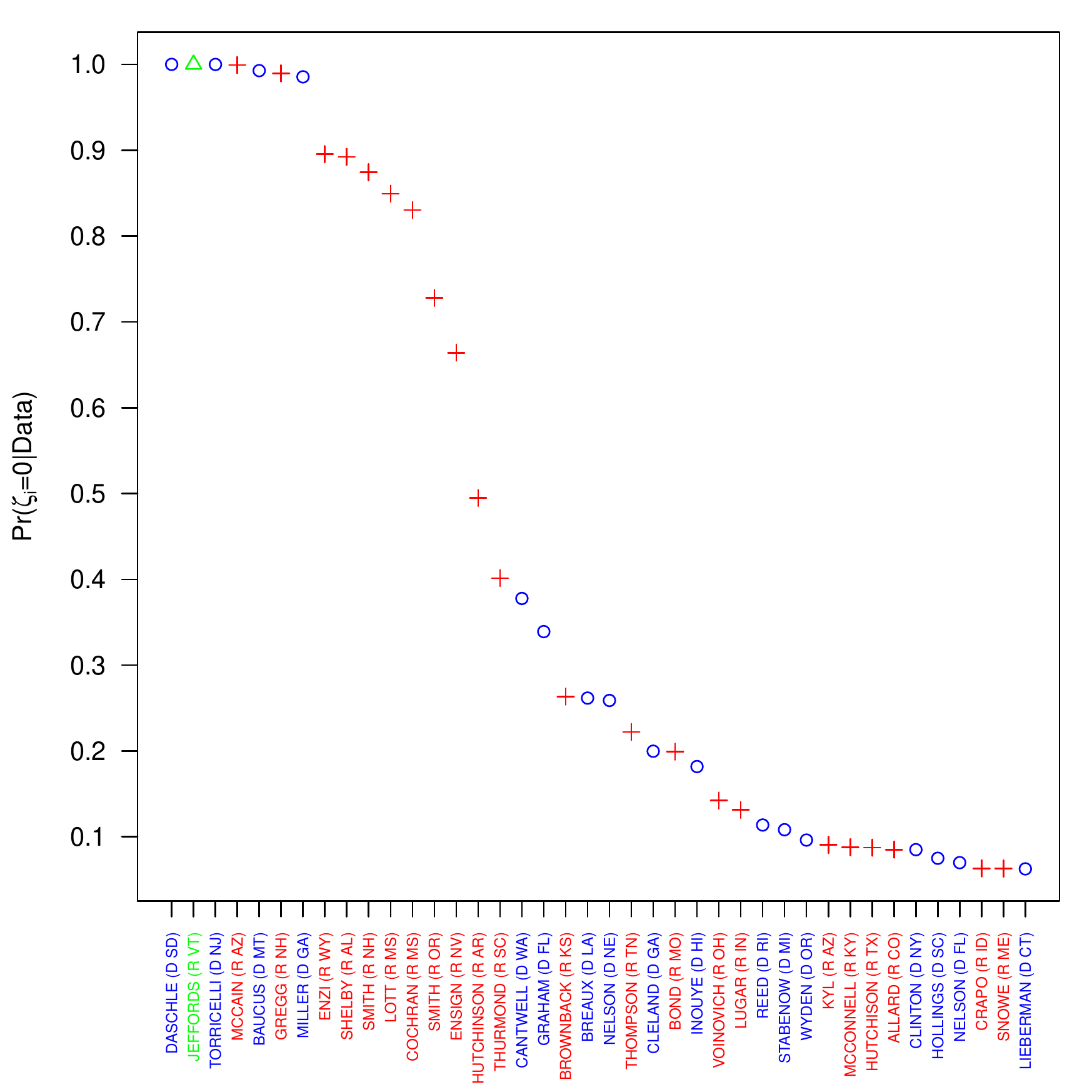}
  \caption{Senators with the 40 largest probabilities of a change in estimated ideal points under our joint unidimensional model with a probit link.  Democratic senators appear as blue circles ({\color{blue} $\circ$}), Republicans appear as red plus signs ({\color{red} $+$}), and party-switcher Jeffords appears as a green triangle ({\color{green} $\triangle$}).  There is at least weak evidence of a change in revealed preferences for 15 legislators, including party leaders Daschle and Lott.}
  \label{fig:promega0}
  \end{center}
\end{figure}

\begin{figure}[!h]
  \begin{center}
  \includegraphics[width=0.8\textwidth]{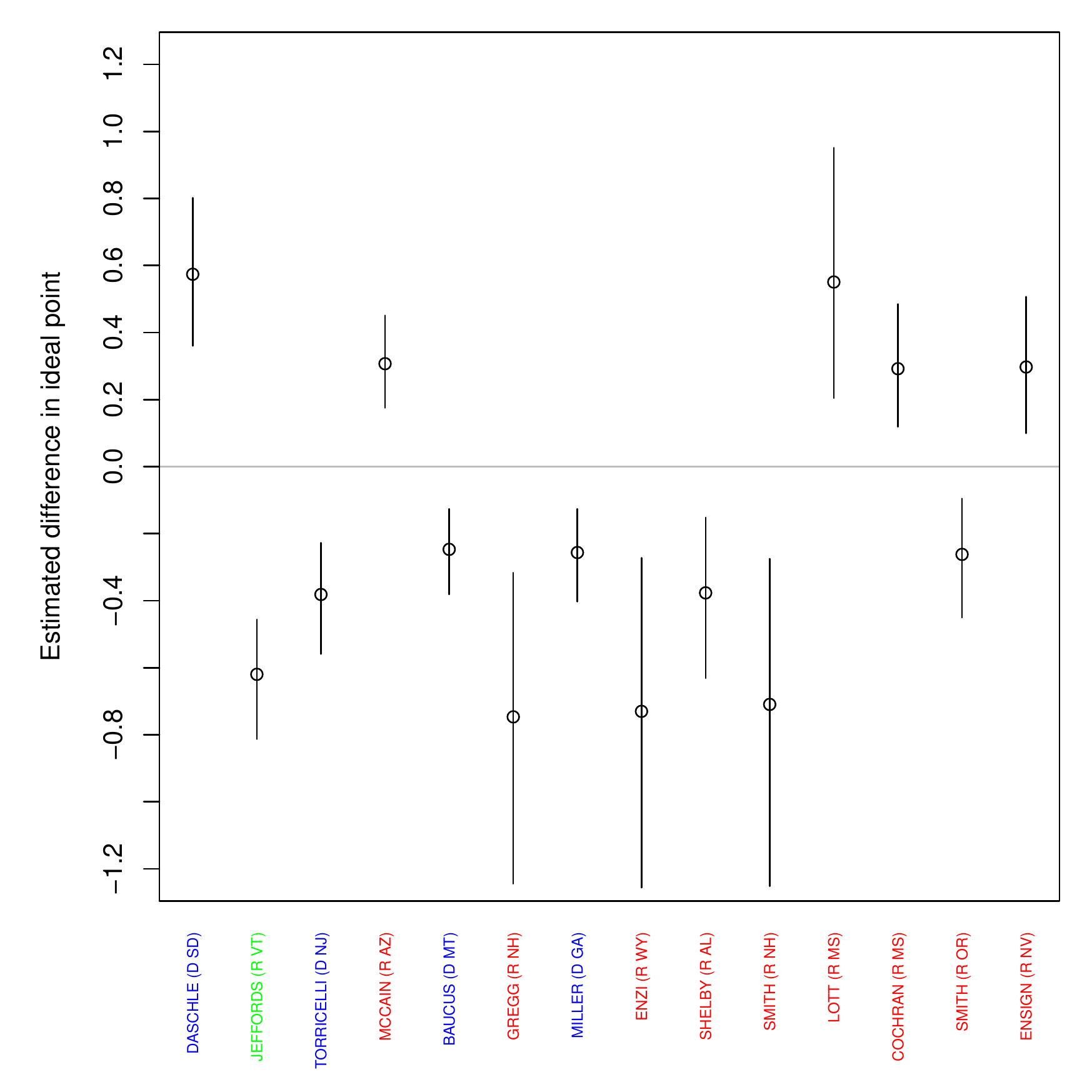}
  \caption{Estimated differences in ideal points (after the switch minus before) and 95\% posterior interval bands for the senators with the largest probabilities of a change in estimated ideal points under our joint unidimensional model with a probit link.  These intervals were constructed conditionally on the a difference being present, i.e., conditional on $\zeta_i = 0$.  Names in red, blue and green correspond to Republican, Democrat and independent senators.  About half of the legislators on each party appear to become more conservative, while the other half appear to become more liberal.}
  \label{fig:topdiffquant}
  \end{center}
\end{figure}

\begin{figure}[!h]
  \begin{center}
    \subfigure[Ideal points]{\includegraphics[width=0.48\textwidth]{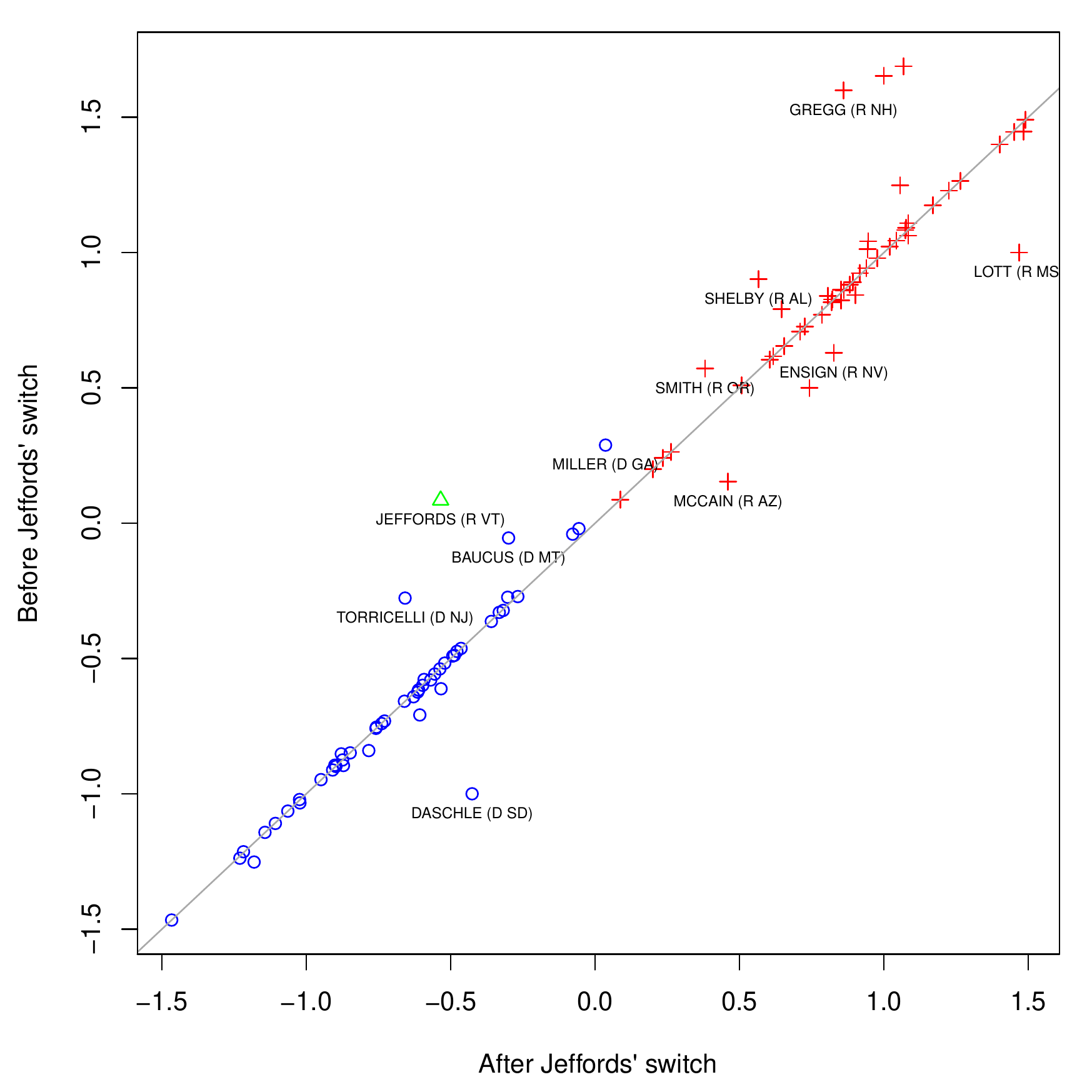}}
    \subfigure[Rank order]{\includegraphics[width=0.48\textwidth]{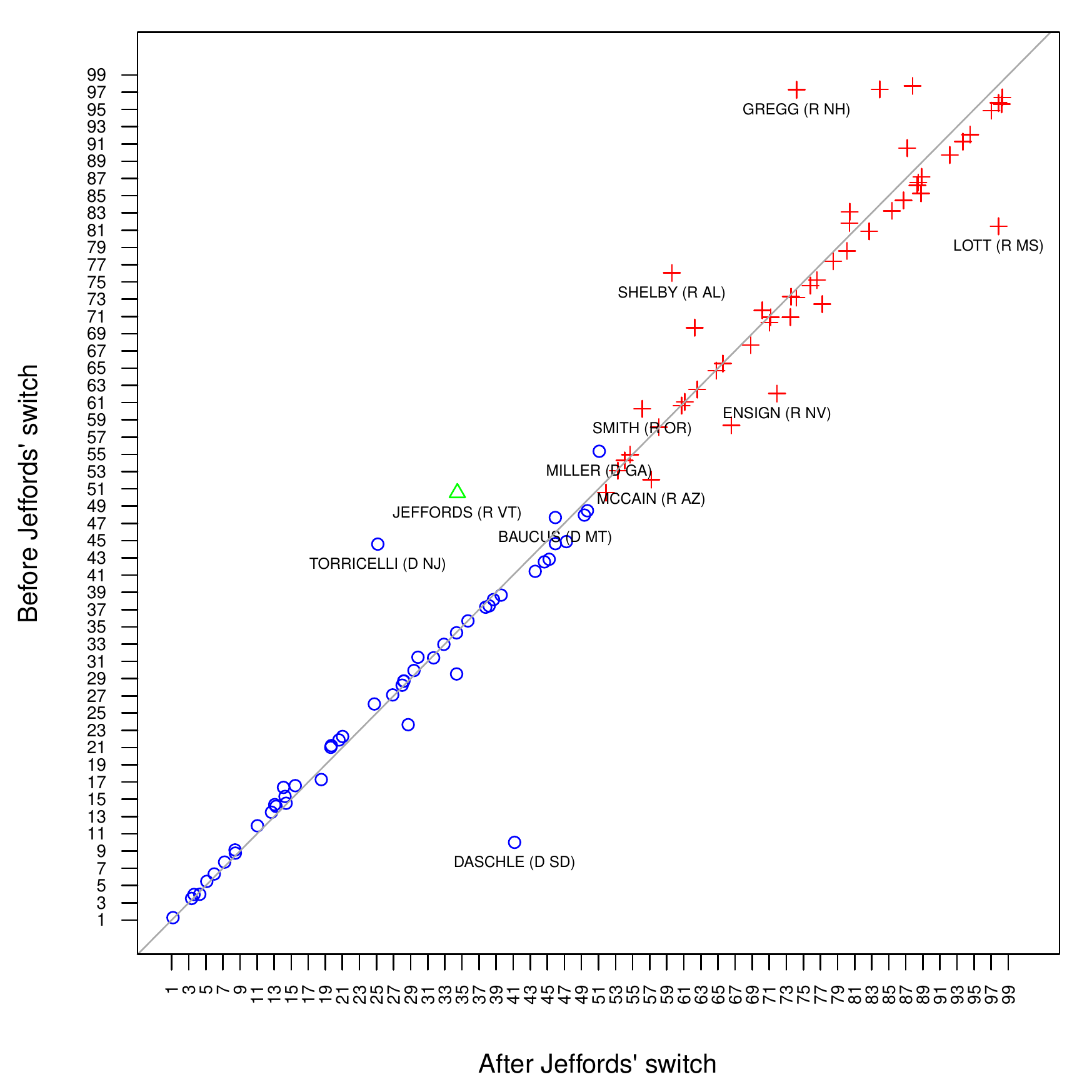}}
  \end{center}
  \caption{Comparison of the posterior mean of the ideal points (left panel) and the rank order of senators (right panel) before and after the switch in Senate control under our joint unidimensional model with a probit link.  Democratic senators appear as blue circles ({\color{blue} $\circ$}), Republicans appear as red plus signs ({\color{red} $+$}), and party-switcher Jeffords appears as a green triangle ({\color{green} $\triangle$}).  Note that the estimates of the change of the order rank of legislators under our joint model are much more conservative than those shown in Figure \ref{fig:independent_ideal_107s}.}
  \label{fig:comparebeta_probit}
\end{figure}

{\color{black} We first present an analysis based on our joint model with a probit link.  We focus on this model first because the probit link was used in previous analyses of the data by \cite{clinton2004statistical} and \cite{roberts2007statistical}.  Figure \ref{fig:promega0} presents the 40 senators our model identify as having the largest posterior probabilities of a change in preferences.  There is evidence of changes in revealed preferences for 14 senators, which could be roughly grouped into three clusters:  a group of seven senators (Torricelli, Daschle, Jeffords, McCain, Miller, Baucus and Gregg) with probabilities above 0.99 (which, following \citealp{kass1995bayes}, we consider very strong evidence of a change in revealed preferences), a second group of five senators (Enzi, Shelby, Smith(NH), Lott, Cochran) with posterior probabilities between 0.8 and 0.9 (which, again following \citealp{kass1995bayes}, we consider strong evidence of a change), and a final cluster of two senators (Smith (OR) and Ensign) with probabilities between 0.65 and 0.75 (showing weak evidence).   In addition to the posterior probabilities, we show in Figure \ref{fig:topdiffquant} symmetric 95\% posterior intervals for the effect of the leadership change on the ideal points of the fourteen senators identified above as presenting changes in revealed preferences (constructed conditionally on the differences being present), and in Figure \ref{fig:comparebeta_probit} the posterior mean of the ideal points (panel (a)) and the posterior mean ranks (panel (b)) for all 100 senators both before and after Jeffords' switch.  From Figure \ref{fig:topdiffquant}, note that the majority of the 14 senators identified as changing their preferences appear to move towards the left; the exceptions are four Republicans (Lott, McCain, Cochran and Ensign) and democratic leader Daschle.  It is also worthwhile noting that the uncertainty associated with senators McCain, Miller and Baucus (who, as we just discussed, are not identified in \cite{clinton2004statistical} as changing their preferences) is relatively low.  From Figure \ref{fig:comparebeta_probit}, note that the difference in ideal points for some of the Senators can be large even if the ranks remain almost unchanged (as is the case with McCain), and vice versa (as is the case with Daschle).  We also note that both the ideal points and the ranks fall very close to the diagonal line, suggesting that our joint model generates much more stable estimates of legislators preferences (at least, when compared with those in Figure \ref{fig:independent_ideal_107s}).

The fact that senator Jeffords' preferences appear to move leftward is intuitively reasonable and consistent with previous work by \cite{nokken2000dynamics} and \cite{nokken2004congressional}.  It supports the theory that changes in party membership of sitting legislators are associated with changes in revealed preferences.  The results are also in line with the analysis in \cite{roberts2007statistical}, who argues that the party leaders should exhibit a change in preferences and (in this case) appear more right-leaning after the Democrats become the majority party.  On the other hand, our results are only partially consistent with those presented in \cite{clinton2004statistical}.  Indeed, although \cite{clinton2004statistical} reported senators Jeffords, Daschle, Lott, Gregg, Smith (NH) and Torricelli as being among the ten legislators with the largest changes in preferences, there are also some striking differences between the two sets of results.  Firstly, senators Kyl, Wellstone and Gramm are identified in \cite{clinton2004statistical} as exhibiting a large change in preferences.  However, our model finds no such evidence; in particular, Wellstone and Gramm do not even make in the list of of the 40 senators with the largest probabilities of a change in preferences. Secondly, three of the senators for which we find evidence of a change in preferences (McCain, Miller, Baucus) are not included in either of \cite{clinton2004statistical} two top-ten lists for changes in preferences.  It is particularly noteworthy that these last three senators are all centrists, while all the ones identified by \cite{clinton2004statistical} but not by our model (Kyl, Wellstone and Gramm) are at the extremes of their respective parties.  The results are also mostly inconsistent with the theory that minority parties tend to stake more extreme positions than their real preferences would indicate \citep{may1973opinion,clausen1987policy,merrill1999power}.  Indeed, in the context of our illustration, this would suggest that both Republicans and Democrats should tend to become more conservative after Jeffords' switch (roughly speaking, Republicans would benefit from staking a more rightward position once they become a minority, while the Democrats lose their incentive to stake a more leftwing position than their preferred one).  However, although this appears to be true for the party leaders (which in their case can be explained by the effect of procedural votes), overall our results provide limited evidence for this phenomenon.}

%
%

\begin{figure}[!h]
  \begin{center}
  \includegraphics[width=0.8\textwidth]{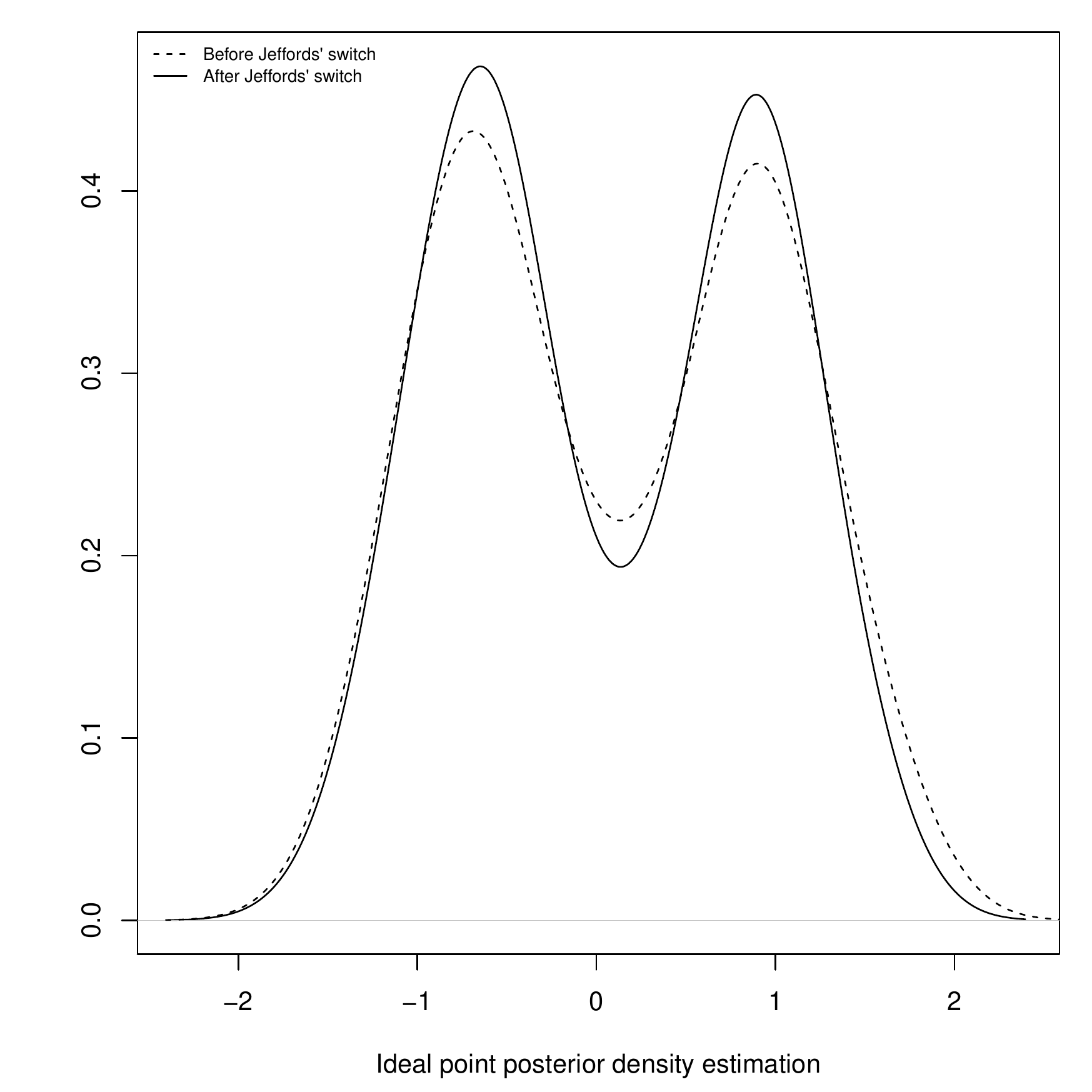}
  \caption{Estimated posterior mean ideal points densities for before and after the majority party switch.  The solid line (\solidrule) is based on  the estimates after Jeffords' switch, while the dashed line (\dashedrule) is based on the estimates before Jeffords' switch.  Note that the two modes seem to become more pronounced (i.e., the parties become more polarized) after senator Jeffords' switch.}
  \label{fig:meanvar}
  \end{center}
\end{figure}

{\color{black}  The previous analysis showed that the revealed preferences displayed by senators were affected by Jefford's switch, but our analysis focused on changes of individual legislators rather than on a summary of the effect of the chamber as a whole.  To address this, Figure \ref{fig:meanvar} presents Gaussian kernel density estimates associated with the posterior means of the ideal points before and after Jeffords' switch.  These kernel density estimates can be interpreted in terms of polarization, a topic that has been widely covered in the political science literature.  Indeed, the relative spread of (intra-party or chamber) estimates is a common way of measuring polarization in political science (e.g., see \citealp{poole1984polarization} and \citealp{mccarty2006polarized}).  Since in this case both sets of ideal points share a common scale, we can extend that analysis to the spread and shape of these two distributions.  In particular, we note that the two humps that can be observed in each density estimate correspond to the two parties represented in the US Congress.  Hence, although both graphs are very similar, it appears as if the parties became somewhat more ideologically homogeneous in terms of their preferences after Jeffords' switch (note that the modes become more pronounced).  There also seems to be a slight tendency of extreme Republicans to become less extreme (as the right tail of the density seems slightly shorter).}

\subsection{Sensitivity analysis}

\begin{figure}[!h]
  \begin{center}
  \includegraphics[width=0.8\textwidth]{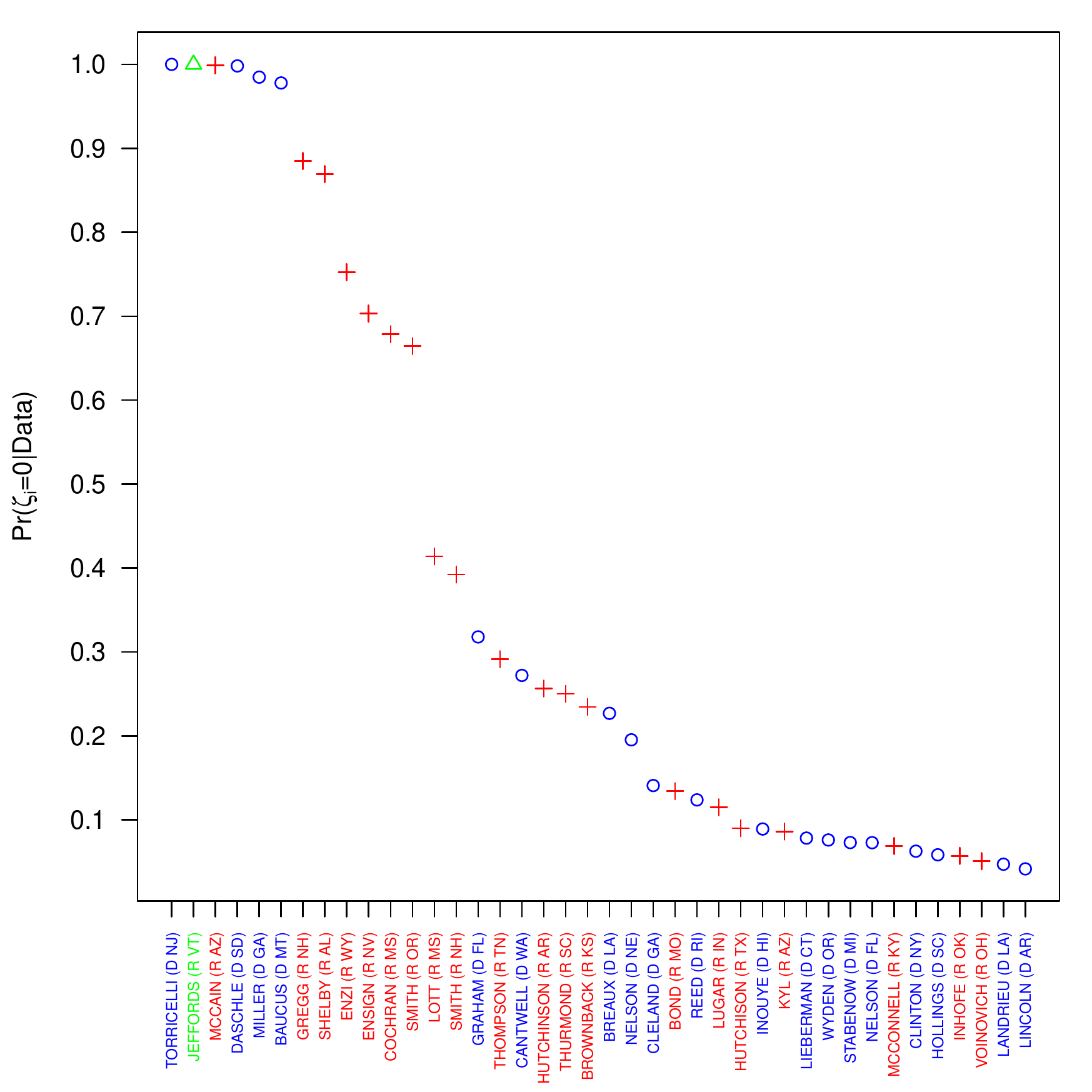}
  \end{center}
  \caption{Senators with the 40 largest probabilities of a change in estimated ideal points under our joint unidimensional model with a logit link.  Democratic senators appear as blue circles ({\color{blue} $\circ$}), Republicans appear as red plus signs ({\color{red} $+$}), and party-switcher Jeffords appears as a green triangle ({\color{green} $\triangle$}).  In this case there is at least weak evidence of a change in revealed preferences for only 12 legislators (note that Republican leader Lott is not among them).}
  \label{fig:promega0_logit}
\end{figure}

To assess the robustness of our analysis to the choice of link functions we reanalyzed the data using a logit link.  As before, Figure \ref{fig:promega0_logit} shows the 40 legislators with the largest posterior probability of a different estimated ideal point.  The conclusions derived from this graph are very similar to those derived from Figure \ref{fig:promega0}.   However, under a logit link the evidence of a change in revealed preferences for senators Lott and Smith (NH) disappears.  We believe that, at least in Republican leader Lott's case, this difference is driven by the fact that assigning a logistic distribution to utility shocks tends to downweight the influence of outliers in the estimates of the ideal points that arise because of procedural rules.  Indeed, recall from Section \ref{se:intro} that U.S.\ Senate rules create incentives for the majority leader to vote against his/her party in certain cloture motions.  These motions are rare, but their influence can make majority leaders appear to be more centrist than they really are \cite{roberts2007statistical}.  What is interesting, though, is that although the conclusion for Republican leader Lott is affected by the link function, the conclusion for Democratic leader Daschle is not.  This is in partial contradiction to the conclusions of \cite{roberts2007statistical} (which are based on comparing ranks of legislators generated by individual probit models), who argues that removing these outlier votes from the analysis eliminates the differences in revealed preferences.
\begin{figure}[!h]
  \begin{center}
  \includegraphics[width=0.8\textwidth]{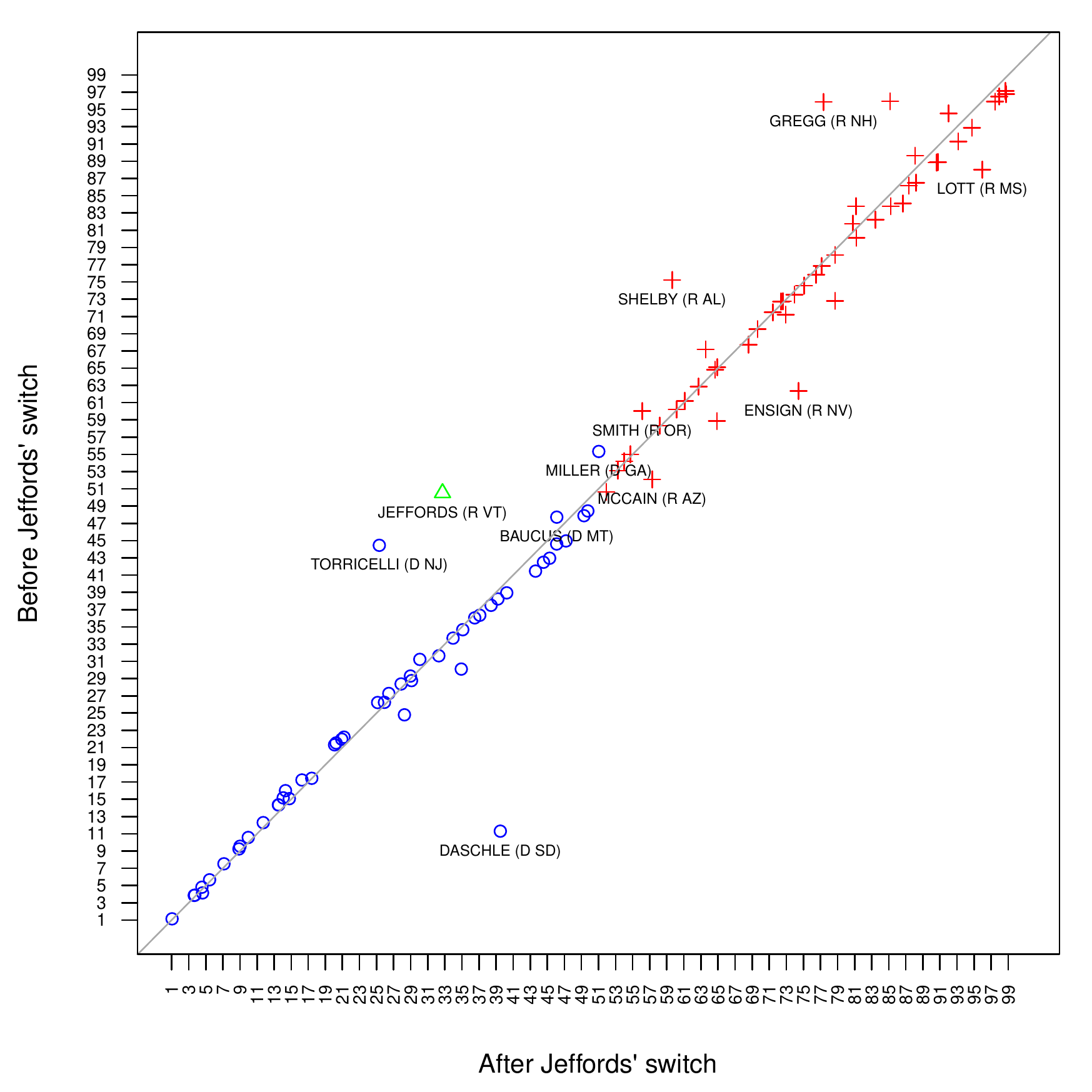}
  \end{center}
  \caption{Posterior mean of the rank order of senators before and after the switch in Senate control under our joint unidimensional model with a logit link.  Democratic senators appear as blue circles ({\color{blue} $\circ$}), Republicans appear as red plus signs ({\color{red} $+$}), and party-switcher Jeffords appears as a green triangle ({\color{green} $\triangle$}).  These estimates are very similar to those shown in Figure \ref{fig:comparebeta_probit}.}
  \label{fig:comparebeta_logit}
\end{figure}

{\color{black} We also investigated the robustness of the results to the prior on the number of bridge legislators.  In addition to our original prior with $a=1$ and $b=9$, we fitted the model using a truncated negative binomial prior with parameters $a=1$ and $b=1$ (which, as we discussed before, implies a uniform prior on the number of bridge legislators).  Figure \ref{fig:altpriorvarpi} presents the equivalent of Figure \ref{fig:promega0} under this prior; note that the conclusions are essentially identical to those obtained under as our original prior.  Similar results were observed when a prior with $a=0.1$ and $b=0.9$ was used (plot not shown).
\begin{figure}[!h]
  \begin{center}
\includegraphics[width=0.8\textwidth]{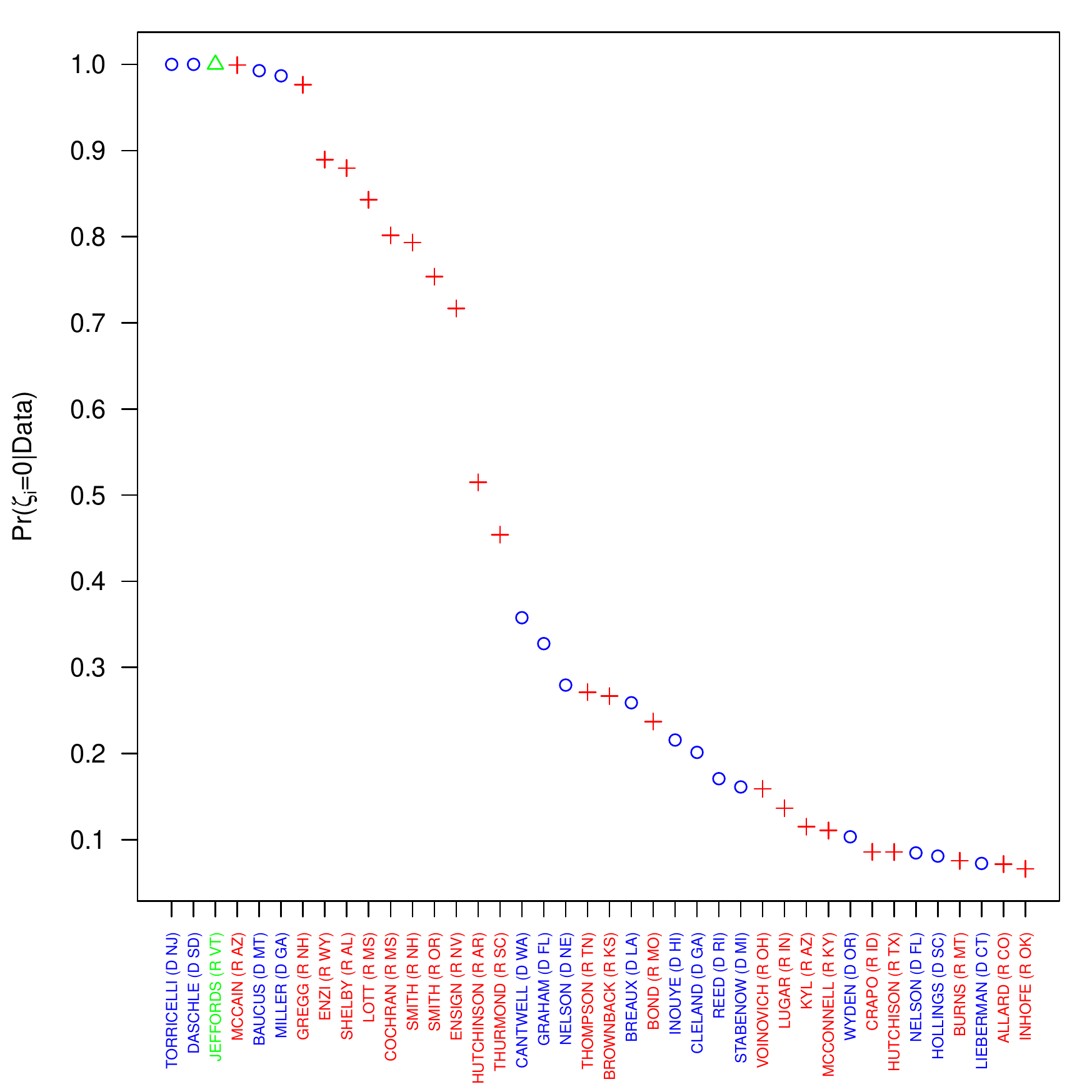}
  \caption{Senators with the 40 largest probabilities of a change in estimated ideal points under our joint unidimensional model with a probit link under the alternate prior $a=1$ and $b=1$.  Democratic senators appear as blue circles ({\color{blue} $\circ$}), Republicans appear as red plus signs ({\color{red} $+$}), and party-switcher Jeffords appears as a green triangle ({\color{green} $\triangle$}).}\label{fig:altpriorvarpi}
  \end{center}
\end{figure}

Another aspect of our sensitivity analysis was an investigation of the robustness of our results to our choice of identifiability constraints.  Recall that we enforced identifiability by fixing the ideal points of the two party leaders (senators Trent Lott and Tom Daschle) to $1$ and $-1$ respectively. Since a good part of our analysis focuses on these two legislators, we also fitted the model by fixing instead the party whips rather than leaders (which in the 107$^{\mbox{th}}$ Senate corresponds to Republican Don Nickles and Democrat Harry Reid) as well as by fixing the two most ``extreme'' legislators (in this case, Republican Hems and Democrat Wellstone). As expected, we saw no difference in the results (plot not shown).}

{\color{black} Finally, we investigated the impact of the priors on $\bfeta$ and $\bfSigma$.  In addition to the original $\normal(\eta \mid 0, 1)$ and $\IGam(\sigma^2 \mid 2,1)$ priors, we also tried overdispersed $\normal(\eta \mid 0, 25)$ and $\IGam(\sigma^2 \mid 2, 25)$ priors.  The results did not change under these priors (plot not shown).}

\subsection{Simulation study and error rates}

In the introduction we argued that one of the advantages of our Bayesian model includes its ability to automatically adjust for multiple comparisons.  Furthermore, we argued that comparisons based on comparing the ideological ranks of the legislators are bound to be less accurate than those based on our model.  To provide some empirical support for this claim we performed a simulation study in which data for 100 legislators and 633 motions (the same numbers as in the 107$^{\mbox{th}}$ Senate) were simulated according to a couple of different scenarios.

\begin{figure}[!h]
  \begin{center}
  \includegraphics[width=0.8\textwidth]{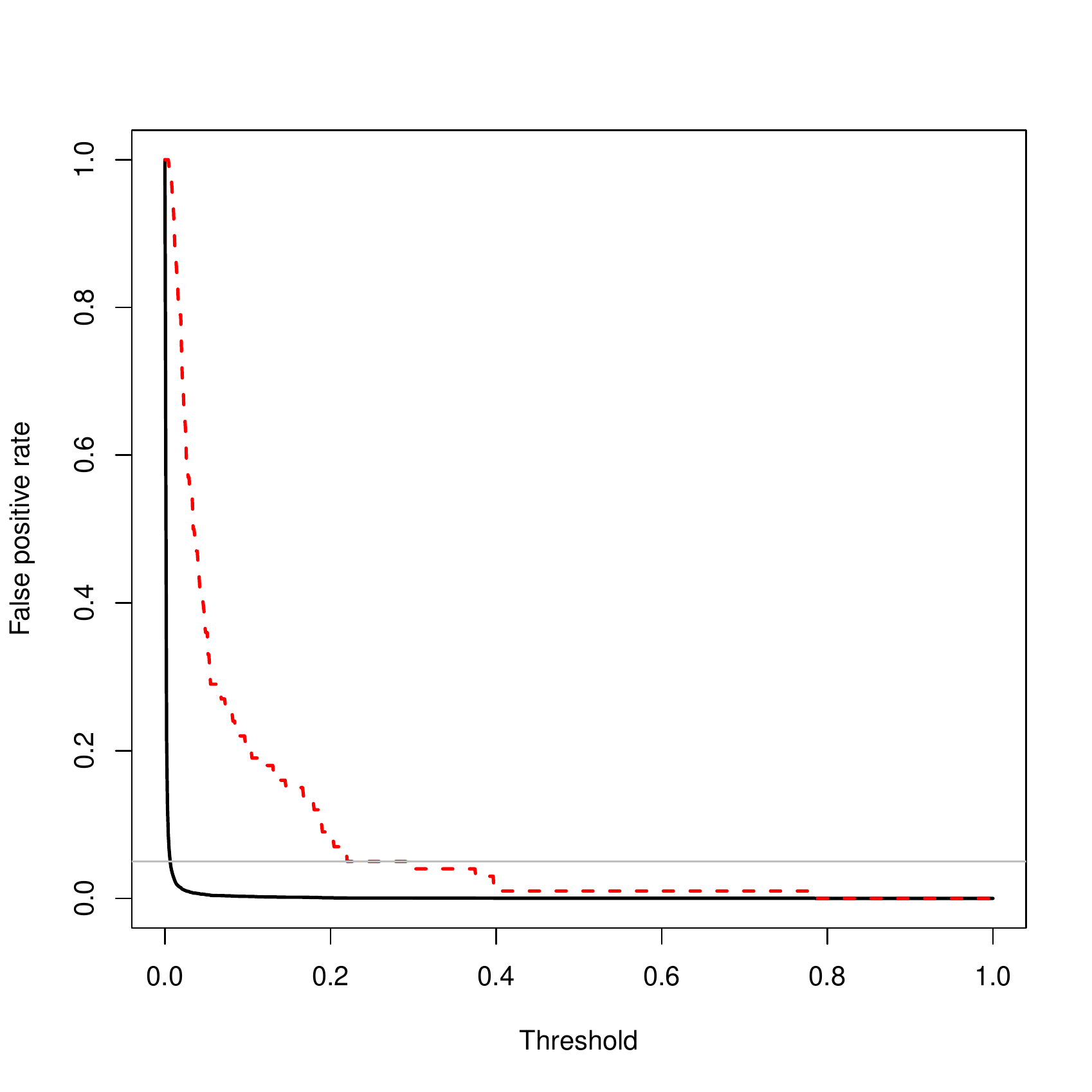}
  \end{center}
  \caption{Mean proportion of false positives (black solid line, \solidrule) and proportion of simulations with at least one false positive (red dashed line, {\color{red} \dashedrule}) for different posterior probability thresholds in our first simulation study.  A false positive rate of 0.05 is shown in grey.  Note that for thresholds above 0.5 both of these error rates are below 1\%.}\label{fig:simstudy1}
\end{figure}

In our first scenario we generated 10 datasets from a logistic unidimensional voting model in which we assumed the same ideal points before and after the majority party switch for all legislators.  We use as true values for the parameters the posterior means obtained by fitting a standard spatial logistic voting model to the full set of roll-call votes from the 107$^{\mbox{th}}$ Senate.  This scenario is used to inquire on the occurrence of false positives, i.e., detecting a difference in ideal points when in truth there is none.  In particular, we study the individual and familywise error rate associated with our testing procedure.  Figure \ref{fig:simstudy1} shows the proportion of simulations in which we identified at least one false positive result (which corresponds to an estimate of the familywise false positive rate) as well as the mean proportion of false positives identified (which provides an estimate of the individual false positive rate on each test) for different posterior probability thresholds.  As expected, the familywise error rate is slightly above the individual error rate.  However, note that for thresholds above 0.5 (which are the most likely to be used in practice) both values are essentially zero.

\begin{figure}[!h]
  \begin{center}
  \subfigure[]{\includegraphics[width=0.49\textwidth]{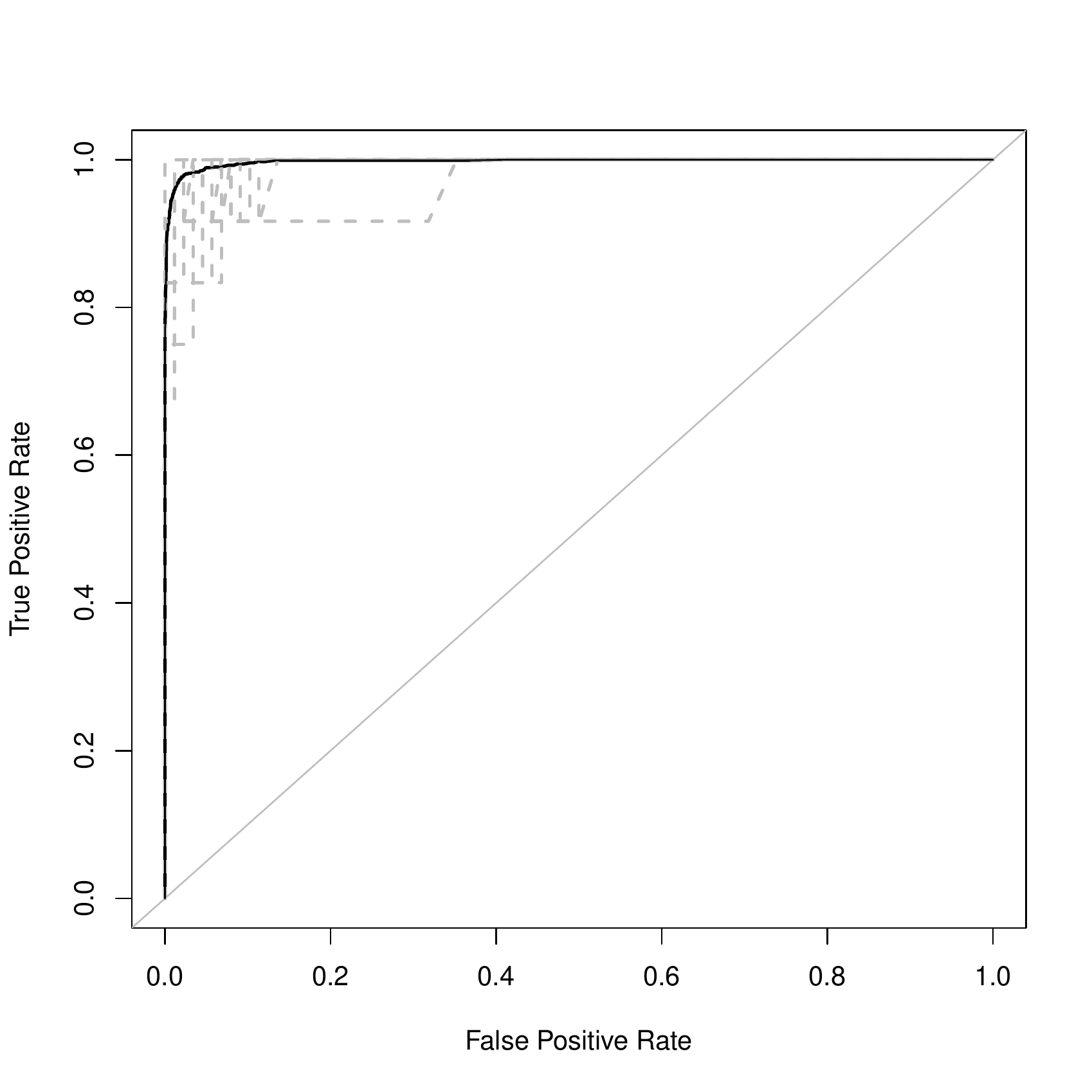}}  
  \subfigure[]{\includegraphics[width=0.49\textwidth]{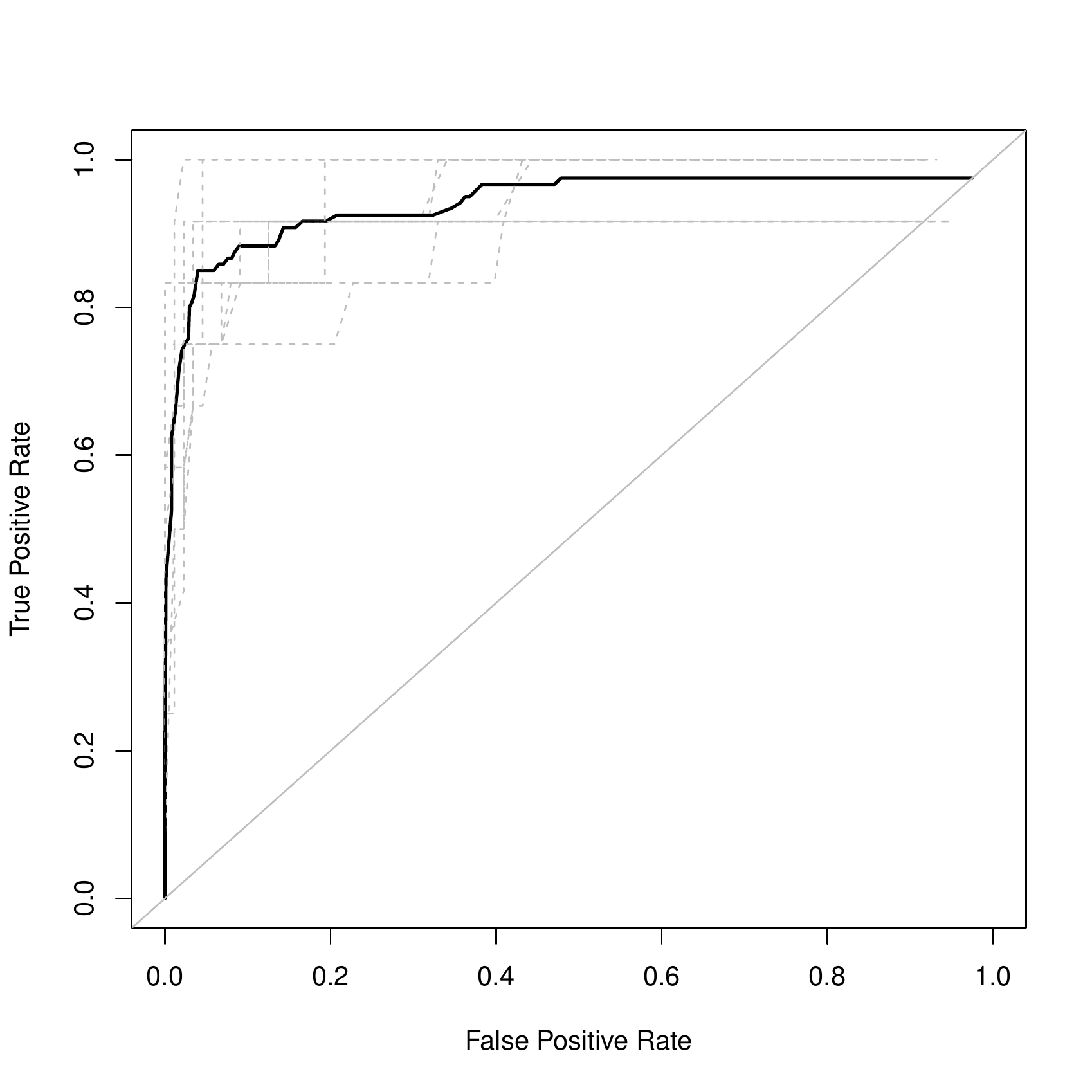}}   
  \end{center}
  \caption{Receiver operating characteristic (ROC) curves for our second simulation study. Panel (a) presents the ROC curves constructed from our method that uses posterior probabilities, while panel (b) shows the ROC curves constructed from the methods used in \cite{clinton2004statistical} and \cite{roberts2007statistical}.  We show both the individual ROC curves associated with each of our simulated datasets (grey dashed lines, {\color{gray} \dashedrule}), as well as the average ROC curve  (black solid line, \solidrule).  The area under the curve (AUC) for the average ROCs curves are 0.98 and 0.86 respectively.}  \label{fig:simstudy2}
\end{figure}

In our second scenario we generated 10 more datasets from a logistic unidimensional voting model in which we assume that the 12 senators we identified in Figure \ref{fig:promega0_logit} are the only ones that exhibit different preferences.  More concretely, the values of the true parameters used to generate these datasets correspond to the posterior means obtained by fitting our joint logit model the data from the 107$^{\mbox{th}}$ Senate assuming that $\zeta_i = 0$ for all senators except Torricelli, Jeffords, McCain, Daschle, Miller, Baucus, Gregg, Shelby, Enzi, Smith (OR), Cochran and Ensign.  In addition to fitting our hierarchical model, we also fit standard logistic models to the roll-calls before and after Jeffords' switch, and compute credible intervals for the difference in rank order in ideological space.  This simulation has two goals.  The first one is to assess both the false positive (detecting a difference in ideal point when in truth there is none) and false negative (failing to detect a difference) rates associated with our procedure. The second is to evaluate the performance of our hierarchical model against a commonly used methodology.  Figure \ref{fig:simstudy2} shows the receiver operating characteristic (ROC) curve associated with each of the two methods.  The curves show that our procedure is highly accurate in detecting true differences whose size is similar to the ones we identified in our analysis of the 107$^{\mbox{th}}$ Senate, and that it outperforms the methodology most commonly used in this context.  In particular, note that the area under the curve (AUC) for the average ROCs curves are 0.98 and 0.86 respectively.

\section{Discussion}\label{se:discussion}

We have presented a statistical model that allows for testing of sharp hypotheses associated with the stability of legislator's preferences inferred from roll call data that fills {\color{black} a methodological shortcoming} in the literature on the analysis of roll-call data.  In particular, our model focuses in the use of zero-inflated priors with carefully constructed hyperpriors that allow us to identify bridge legislators that connect the arbitrary ideological scales associated with different groups of motions and make them comparable.

In addition to our methodological contribution, we also present a detailed analysis of the voting record of the 107$^{\mbox{th}}$ U.S.\ Senate, which saw a major reorganization (which included a switch in agenda-setting powers from the Republican to the Democratic party) in the middle of the session due to a change in party affiliation of senator James M.\ Jeffords.  {\color{black} Because the time period under consideration is relative short and Jeffords' switch is as close as to a surprise shock as we are likely to see, we have interpreted the observed changes in revealed preferences as being causally related to the switch.  In particular, we have argued that our analysis provides empirical evidence for the effect of party membership on revealed preferences (we find strong evidence that senator Jeffords' preferences appear to be different after he leaves the Republican party), but no evidence for the effect of majority status on legislators preferences (as most legislators who had their preferences affected by the reorganization tend to become liberal rather than conservative).  However, we emphasize that, as with other analyses, care needs to be excersized when making causal attributions for the differences identified by our model.}

{\color{black} Our analysis of the 107$^{\mbox{th}}$ U.S.\ Senate also highlights the pitfalls associated with assessing the stability of legislators' revealed preferences using ad-hoc adjustments to the latent ideological scales (as in \citealp{clinton2004statistical}) or by focusing on the rank order of legislators in the ideological space (as in \citealp{roberts2007statistical}).  It also highlights the potential impact of different link functions on the analysis.  Indeed, although both the probit and logit links tend to provide similar results, the fact that the results of the analysis for Republican leader Trent Lott are affected by the choice of link function combined with our knowledge that procedural rules provide specific incentives for party leaders to deviate from their preferred behavior, suggest that analyses of the stability of preferences should be conducted using links such as the logit, allowing the model to be more robust to outliers.}

{\color{black} Finally, we note that two interesting extensions of our model involve comparing more than two groups of motions (e.g. comparing types of motions such as procedural, amendment, final passage, etc., see for example \citealp{jess14-two}) and incorporating point mass priors on dynamic models such as those in \cite{martin2002dynamic}.  These extensions will be pursued elsewhere.}

\bibliographystyle{apalike}

\bibliography{advancementbib}

\end{document}


\maketitle
This supplemental materials present a detailed description of the full conditional distributions associated with the Markov chain Monte Carlo algorithms used to fit the models described in ``Assessing differences in legislators' revealed preferences:  A case study on the 107$^{\mbox{th}}$ U.S. Senate'' for the case $K=1$ (i.e., a unidimensional policy space).

\section{Probit algorithm}

\begin{enumerate}

\item Sample the parameters associated with the policy space.
\begin{enumerate}
\item For $j = 1,...,J$ sample $\alpha^*_j|\cdots \sim \hat{\omega}_\alpha \delta_0(\alpha_j) + (1-\hat{\omega}_{\alpha}) N(\hat{\rho}_\alpha,\hat{\kappa}_\alpha^2)$ where
\begin{align*}
\hat{\rho}_\alpha &= \hat{\kappa}_\alpha^2 \left( \sum\limits_{i=1}^I (z_{ij}-\mu_j)\beta_{i,\gamma_j} \right), &
\hat{\kappa}_\alpha^2 &= \left( \sum\limits_{i=1}^I \beta_{i,\gamma_j}^2 + \frac{1}{\kappa_\alpha^2} \right)^{-1}  ,  &
\hat{\omega}_\alpha &= \frac{\omega_\alpha\kappa_\alpha^2}{\omega_\alpha\kappa_\alpha^2 + (1-\omega_\alpha)\hat{\kappa}_\alpha^2\exp\left( \frac{\hat{\rho}_\alpha^2}{2\hat{\kappa}_\alpha^2} \right)}  .
\end{align*}

\item For $j = 1,...,J$ sample $\mu^*_j|\cdots \sim N(\hat{\rho}_\mu,\hat{\kappa}_\mu^2)$ where
\begin{align*}
&\hat{\rho}_\mu = \hat{\kappa}_\mu^2 \left( \sum\limits_{i=1}^I (z_{ij}-\alpha_j\beta_{i,\gamma_j}) + \frac{\rho_\mu}{\kappa_\mu^2} \right), &\hat{\kappa}_\mu^2 = \left( I + \frac{1}{\kappa_\mu^2} \right)^{-1}  .
\end{align*}

\item For $i=1,...,I$ sample $\zeta_i|\cdots \sim Bern(p)$ where
\[ p =
  \begin{cases}
    1       & \quad \text{if } \sum\limits_{i=1}^I \zeta_i = K+1 \\
    \frac{p_1}{p_1+p_0}  & \quad \text{if } \sum\limits_{i=1}^I \zeta_i > K+1 \\
  \end{cases}
\] ,
\begin{multline*}
p_1 \propto \Gamma\left( a+1+\sum\limits_{j\ne i} \zeta_j \right) \Gamma\left( b+I-1-\sum\limits_{j\ne i}\zeta_j \right) \left( \frac{1}{\sigma} \right) \left( \sum\limits_{j=1}^J \alpha_j^2 + \frac{1}{\sigma^2} \right)^{-1/2} \\
\times \exp\left\{  -\frac{1}{2}\left[ \sum\limits_{j=1}^J(z_{ij}-\mu_j)^2 + \frac{\eta^2}{\sigma^2} -\left( \sum\limits_{j=1}^J (z_{ij}-\mu_j)\alpha_j + \frac{\eta}{\sigma^2} \right)^2\left( \sum\limits_{j=1}^J\alpha_j^2 + \frac{1}{\sigma^2} \right)^{-1} \right]\right\}, \\
\end{multline*}
and
\begin{multline*}
p_0 \propto \Gamma\left( a+\sum\limits_{j\ne i} \zeta_j \right) \Gamma\left( b+I-\sum\limits_{j\ne i}\zeta_j \right) \left( \frac{1}{\sigma^2} \right) \left( \sum\limits_{j:\gamma_j=0} \alpha_j^2 + \frac{1}{\sigma^2} \right)^{-1/2} \left( \sum\limits_{j:\gamma_j=1} \alpha_j^2 + \frac{1}{\sigma^2} \right)^{-1/2} \\
\times \exp\Bigg\{-\frac{1}{2}\Bigg[ \sum\limits_{j=1}^J(z_{ij}-\mu_j)^2 + \frac{2\eta^2}{\sigma^2} -\left( \sum\limits_{j:\gamma_j=0}(z_{ij}-\mu_j)\alpha_j + \frac{\eta}{\sigma^2} \right)^2\left( \sum\limits_{j:\gamma_j=0}\alpha_j^2 + \frac{1}{\sigma^2} \right)^{-1} \\
 - \left( \sum\limits_{j:\gamma_j=1}(z_{ij}-\mu_j)\alpha_j + \frac{\eta}{\sigma^2} \right)^2\left( \sum\limits_{j:\gamma_j=1}\alpha_j^2 + \frac{1}{\sigma^2} \right)^{-1} \Bigg]\Bigg\}  .
\end{multline*}

\item  For $i = 1,...,I$ sample $\beta^*_i \mid \zeta_i=1,\cdots \sim N(\hat{\rho}_\beta,\hat{\kappa}^2_\beta)$ where
\begin{align*}
\hat{\rho}_\beta &= \hat{\kappa}_\beta^2 \left( \sum\limits_{j=1}^J (z_{ij}-\mu_j)\alpha_j + \frac{\eta}{\sigma^2} \right), &  \hat{\kappa}_\beta^2 &= \left( \sum\limits_{j=1}^J \alpha_j^2 + \frac{1}{\sigma^2} \right)^{-1}  .
\end{align*}
and for $l=0,1$ sample $\beta^*_{il}|\zeta_i=0 \sim N(\hat{\rho}_{\beta_l}, \hat{\kappa}^2_{\beta_l}) $
\begin{align*}
&\hat{\rho}_{\beta_l} = \hat{\kappa}_{\beta_l}^2 \left( \sum\limits_{j:\gamma_j=l} (z_{ij}-\mu_j)\alpha_j + \frac{\eta}{\sigma^2} \right), &\hat{\kappa}_{\beta_l}^2 = \left( \sum\limits_{j:\gamma_j=l} \alpha_j^2 + \frac{1}{\sigma^2} \right)^{-1}  .
\end{align*}
\end{enumerate}

\item Enforce identifiability constraint. In the case of a unidimensional setting we often fix the position of the party leaders so that $\beta_{DemLeader,0} = -1$ and $\beta_{RepLeader,0} = 1$, so that the transformed ideal points correspond to 
\begin{align*}
\beta_{i,l} &= \frac{2\beta^*_{i,l}-\beta^*_{DemLeader,0} -\beta^*_{RepLeader,0}}{\beta^*_{DemLeader,0} -\beta^*_{RepLeader,0}}  ,
\end{align*}
while the motion-specific parameters become
\begin{align*}
\mu_j &= \mu^*_j + \alpha^*_j \left(\frac{\beta^*_{DemLeader,0} +\beta^*_{RepLeader,0}}{2}\right)  , &
\alpha_j &= \alpha^*_j \left(\frac{\beta^*_{DemLeader,0} - \beta^*_{RepLeader,0}}{2}\right)  , &
\end{align*}
where $\mu^*_j$, $\alpha^*_j$, and $\beta^*_{i,l}$ represent the unconstrained parameter values.

\item For $i=1,...I$ and $j=1,...,J$ sample
\[ z_{ij} \sim
  \begin{cases}
    N(\mu_j+\alpha_j\beta_{i,\gamma_j}) \mathbbm{1}_{z_{ij} \ge 0}       & \quad \text{if } y_{ij} = 1 \\
    N(\mu_j+\alpha_j\beta_{i,\gamma_j}) \mathbbm{1}_{z_{ij} < 0}  & \quad \text{if } y_{ij} = 0 \\
  \end{cases}  .
\]

\item Sample hyperparameters.
\begin{enumerate}
\item Sample $\omega_\alpha \sim Beta\left( a + \sum\limits_{j:\alpha_j=0}1, b + J - \sum\limits_{j:\alpha_j=0}1 \right)$ .
\item Sample $\kappa_\alpha^2 \sim IG\left( 2 + J/2, 1 + \sum\limits_{j=1}^J \frac{\alpha_j^2}{2} \right)$ .
\item Sample $\rho_\mu \sim N \left( \left( \sum\limits_{j=1}^J \frac{\mu_j}{\kappa_\mu^2} \right) \left( 1 + J/\kappa_\mu^2 \right)^{-1} ,\left( 1 + J/\kappa_\mu^2 \right)^{-1} \right)$ .
\item Sample $\kappa_\mu^2 \sim IG\left( 2 + J/2, 1 + \sum\limits_{j=1}^J \frac{(\mu_j - \rho_\mu)^2}{2} \right)$ .
\item Sample $\eta \sim N \left( \left( \frac{\sum\limits_{i:\zeta_i=1}\beta_{i,0}+ \sum\limits_{i:\zeta_i=0}(\beta_{i,0}+\beta_{i,1})}{\sigma^2} \right) \left( 1 + \frac{I + \sum\limits_{i=1}^I \zeta_i}{\sigma^2}  \right)^{-1} ,\left( 1 + \frac{I + \sum\limits_{i=1}^I \zeta_i}{\sigma^2} \right)^{-1} \right)$ .
\item Sample $\sigma^2 \sim IG\left( 2 + \frac{I+\sum\limits_{i=1}^I\zeta_i}{2}, 1 + I + \frac{\sum\limits_{i:\zeta_i=1} (\beta_{i,0} - \eta)^2 + \sum\limits_{i: \zeta_i=0}\left( (\beta_{i,0}-\eta)^2 + (\beta_{i,1}-\eta)^2 \right)}{2} \right)$ .
\end{enumerate}
where $IG(a,b)$ denotes the Inverse-Gamma distribution with mean $\frac{b}{a-1}$.

\item Sample missing observations. If $y_{ij}$ is missing, sample $y_{ij} \sim Bern(\Phi(\mu_j+\alpha_j\beta_{i,\gamma_j}))$ where $\Phi$ denotes the Normal distribution function.

\end{enumerate}

\section{Logit algorithm}

\begin{enumerate}

\item Sample the parameters associated with the policy space.
\begin{enumerate}
\item For $j = 1,...,J$ sample $\alpha^*_j|\cdots \sim \hat{\omega}_\alpha \delta_0(\alpha_j) + (1-\hat{\omega}_{\alpha}) N(\hat{\rho}_\alpha,\hat{\kappa}_\alpha^2)$ where
\begin{align*}
\hat{\rho}_\alpha &= \hat{\kappa}_\alpha^2 \left( \sum\limits_{i=1}^I \left(y_{ij}-\frac{1}{2}\right)\beta_{i,\gamma_j} - \sum\limits_{i=1}^I \mu_j\beta_{i,\gamma_j}w_{ij} \right)  , &
\hat{\kappa}_\alpha^2 &= \left( \sum\limits_{i=1}^I \beta_{i,\gamma_j}^2w_{ij} + 1/\kappa_\alpha^2 \right)^{-1}  , 
\end{align*}
and
\begin{align*}
\hat{\omega}_\alpha &= \frac{\omega_\alpha\kappa_\alpha^2}{\omega_\alpha\kappa_\alpha^2 + (1-\omega_\alpha)\hat{\kappa}_\alpha^2\exp\left( \frac{\hat{\rho}_\alpha^2}{2\hat{\kappa}_\alpha^2} \right)}  .
\end{align*}

\item For $j = 1,...,J$ sample $\mu^*_j|\cdots \sim N(\hat{\rho}_\mu,\hat{\kappa}_\mu^2)$ where
\begin{align*}
&\hat{\rho}_\mu = \hat{\kappa}_\mu^2 \left( \sum\limits_{i=1}^I \left(y_{ij}-\frac{1}{2}\right) - \sum\limits_{i=1}^I\alpha_j\beta_{i,\gamma_j}w_{ij} + \frac{\rho_\mu}{\kappa_\mu^2}\right), &\hat{\kappa}_\mu^2 = \left(  \sum\limits_{i=1}^I w_{ij} + \frac{1}{\kappa_\mu^2} \right)^{-1} .
\end{align*}

\item For $i=1,...,I$ sample $\zeta_i|\cdots \sim Bern(p)$ where
\[ p =
  \begin{cases}
    1       & \quad \text{if } \sum\limits_{i=1}^I \zeta_i = K+1 \\
    \frac{p_1}{p_1+p_0}  & \quad \text{if } \sum\limits_{i=1}^I \zeta_i > K+1 \\
  \end{cases}
\]  
\begin{multline*}
p_1 \propto \Gamma \left( a+1+\sum\limits_{j\ne i} \zeta_j \right) \Gamma\left( b+I-1-\sum\limits_{j\ne i}\zeta_j \right) \left( \frac{1}{\sigma} \right) \left( \sum\limits_{j=1}^J \alpha_j^2w_{ij} + \frac{1}{\sigma^2} \right)^{-1/2} \\
\times \exp\Bigg\{ -\frac{1}{2}\Bigg[ \sum\limits_{j=1}^J2\left(y_{ij}-\frac{1}{2}\right)\mu_j + \sum\limits_{j=1}^J\mu_j^2w_{ij} + \frac{\eta^2}{\sigma^2} \\
-\left( \sum\limits_{j=1}^J \left(y_{ij}-\frac{1}{2}\right)\alpha_j  - \sum\limits_{j=1}^J\mu_j\alpha_jw_{ij} + \frac{\eta}{\sigma^2} \right)^2\left( \sum\limits_{j=1}^J\alpha_j^2w_{ij} + \frac{1}{\sigma^2} \right)^{-1} \Bigg]\Bigg\},
\end{multline*}
and
\begin{multline*}
p_0 \propto \Gamma \left( a+\sum\limits_{j\ne i} \zeta_j \right) \Gamma\left( b+I-\sum\limits_{j\ne i}\zeta_j \right) \left( \frac{1}{\sigma^2} \right) \left( \sum\limits_{j:\gamma_j=0} \alpha_j^2w_{ij} + \frac{1}{\sigma^2} \right)^{-1/2}  \left( \sum\limits_{j:\gamma_j=1} \alpha_j^2w_{ij} + \frac{1}{\sigma^2} \right)^{-1/2} \\
\times \exp\Bigg\{-\frac{1}{2}\Bigg[ \sum\limits_{j=1}^J2\left(y_{ij}-\frac{1}{2}\right)\mu_j + \sum\limits_{j=1}^J\mu_j^2w_{ij} + \frac{2\eta^2}{\sigma^2} \\
-\left( \sum\limits_{j:\gamma_j=0} \left(y_{ij}-\frac{1}{2}\right)\alpha_j  - \sum\limits_{j:\gamma_j=0}\mu_j\alpha_jw_{ij} + \frac{\eta}{\sigma^2} \right)^2\left( \sum\limits_{j:\gamma_j=0}\alpha_j^2w_{ij} + \frac{1}{\sigma^2} \right)^{-1} \\
-\left( \sum\limits_{j:\gamma_j=1} \left(y_{ij}-\frac{1}{2}\right)\alpha_j  - \sum\limits_{j:\gamma_j=1}\mu_j\alpha_jw_{ij} + \frac{\eta}{\sigma^2} \right)^2\left( \sum\limits_{j:\gamma_j=1}\alpha_j^2w_{ij} + \frac{1}{\sigma^2} \right)^{-1}\Bigg]\Bigg\}  .
\end{multline*}

\item  For $i = 1,...,I$ sample $\beta^*_i|\zeta_i=1,\cdots \sim N(\hat{\rho}_\beta,\hat{\kappa}^2_\beta)$ where
\begin{align*}
&\hat{\rho}_\beta = \hat{\kappa}_\beta^2 \left( \sum\limits_{j=1}^J \left(y_{ij}-\frac{1}{2}\right)\alpha_j - \sum\limits_{j=1}^J\mu_j\alpha_jw_{ij} + \frac{\eta}{\sigma^2} \right), &\hat{\kappa}_\beta^2 = \left( \sum\limits_{j=1}^J \alpha_j^2w_{ij} + \frac{1}{\sigma^2} \right)^{-1}  .
\end{align*}
and for $l=0,1$ sample $\beta^*_{il}|\zeta_i=0 \sim N(\hat{\rho}_{\beta_l}, \hat{\kappa}^2_{\beta_l}) $
\begin{align*}
&\hat{\rho}_{\beta_l} = \hat{\kappa}_{\beta_l}^2 \left( \sum\limits_{j:\gamma_j=l} \left(y_{ij}-\frac{1}{2}\right)\alpha_j - \sum\limits_{j:\gamma_j=l}\mu_j\alpha_jw_{ij} + \frac{\eta}{\sigma^2} \right), &\hat{\kappa}_{\beta_l}^2 = \left( \sum\limits_{j:\gamma_j=l} \alpha_j^2w_{ij} + \frac{1}{\sigma^2} \right)^{-1}  .
\end{align*}
\end{enumerate}

\item Enforce identifiability constraint. In the case of a unidimensional setting we often fix the position of the party leaders so that $\beta_{DemLeader,0} = -1$ and $\beta_{RepLeader,0} = 1$, so that the transformed ideal points correspond to 
\begin{align*}
\beta_{i,l} &= \frac{2\beta^*_{i,l}-\beta^*_{DemLeader,0} -\beta^*_{RepLeader,0}}{\beta^*_{DemLeader,0} -\beta^*_{RepLeader,0}}  ,
\end{align*}
while the motion-specific parameters become
\begin{align*}
\mu_j &= \mu^*_j + \alpha^*_j \left(\frac{\beta^*_{DemLeader,0} +\beta^*_{RepLeader,0}}{2}\right)  , &
\alpha_j &= \alpha^*_j \left(\frac{\beta^*_{DemLeader,0} - \beta^*_{RepLeader,0}}{2}\right)  , &
\end{align*}
where $\mu^*_j$, $\alpha^*_j$, and $\beta^*_{i,l}$ represent the unconstrained parameter values.

\item For $i=1,...I$ and $j=1,...,J$ sample $w_{ij} \sim Polya-Gamma(1,\mu_j+\alpha_j\beta_{i,\gamma_j})$.

\item Sample hyperparameters.
\begin{enumerate}
\item Sample $\kappa_\alpha^2 \sim IG\left( 2 + J/2, 1 + \sum\limits_{j=1}^J \frac{\alpha_j^2}{2} \right)$.
\item Sample $\rho_\mu \sim N \left( \left( \sum\limits_{j=1}^J \frac{\mu_j}{\kappa_\mu^2} \right) \left( 1 + J/\kappa_\mu^2 \right)^{-1} ,\left( 1 + J/\kappa_\mu^2 \right)^{-1} \right)$.
\item Sample $\kappa_\mu^2 \sim IG\left( 2 + J/2, 1 + \sum\limits_{j=1}^J \frac{(\mu_j - \rho_\mu)^2}{2} \right)$.
\item Sample $\eta \sim N \left( \left( \frac{\sum\limits_{i:\zeta_i=1}\beta_{i,0}+ \sum\limits_{i:\zeta_i=0}(\beta_{i,0}+\beta_{i,1})}{\sigma^2} \right) \left( 1 + I + \frac{\sum\limits_{i=1}^I \zeta_i}{\sigma^2}  \right)^{-1} ,\left( 1 + I + \frac{\sum\limits_{i=1}^I \zeta_i}{\sigma^2} \right)^{-1} \right)$.
\item Sample $\sigma^2 \sim IG\left( 2 + \frac{I+\sum\limits_{i=1}^I\zeta_i}{2}, 1 + I + \frac{\sum\limits_{i:\zeta)_i=1} (\beta_{i,0} - \eta)^2 + \sum\limits_{i:\zeta_i=0}\left( (\beta_{i,0}-\eta)^2 + (\beta_{i,1}-\eta)^2 \right)}{2} \right)$.

\end{enumerate}
where $IG(a,b)$ denotes the Inverse-Gamma distribution with mean $\frac{b}{a-1}$.

\item Sample missing observations. If $y_{ij}$ is missing, sample $y_{ij} \sim Bern(G(\mu_j+\alpha_j\beta_{i,\gamma_j}))$, where $G$ denotes the Logistic distribution function, $G(x) = \left(1 + \exp\{-x\}\right)$.
\end{enumerate}